\documentclass[]{aa}  

\usepackage{graphicx}
\usepackage{txfonts}
\usepackage{xspace}
\usepackage{booktabs}
\usepackage{float}
\usepackage{soul}

\usepackage[table, dvipsnames,svgnames,x11names]{xcolor}
\usepackage[colorlinks=true, linktocpage, linkcolor={blue!60!black}, citecolor={blue!60!black}, urlcolor={blue!60!black}]{hyperref}
\usepackage{subcaption}
\usepackage{tikz}

\begin{document}

   \title{The conceptual design of the 50-meter Atacama Large Aperture Submillimeter Telescope (AtLAST)}
   \titlerunning{The AtLAST conceptual design}
   \authorrunning{The AtLAST Telescope Design Team}
   
   \author{Tony Mroczkowski\inst{1}\thanks{contact \email{tonym@eso.org}},
          Patricio A. Gallardo\inst{2},
          Martin Timpe\inst{3}, 
          Aleksej Kiselev\inst{3}, 
          Manuel Groh\inst{3},
          Hans Kaercher\inst{4}, 
          Matthias Reichert\inst{3}, 
          Claudia Cicone\inst{5},
          Roberto Puddu\inst{6},
          Pierre Dubois-dit-Bonclaude\inst{3},
          Daniel Bok\inst{3}, 
          Erik Dahl\inst{3},
          Mike Macintosh\inst{7},
          Simon Dicker\inst{8},
          Isabelle Viole\inst{9},
          Sabrina Sartori\inst{9},
          Guillermo Andr\'es Valenzuela Venegas\inst{9},
          Marianne Zeyringer\inst{9},
          Michael Niemack\inst{10,11}, 
          Sergio Poppi\inst{12}, 
          Rodrigo Olguin\inst{13},
          Evanthia Hatziminaoglou\inst{1,14,15},
          Carlos De Breuck\inst{1},
          Pamela Klaassen\inst{7},
          Francisco Miguel Montenegro-Montes\inst{1,16},
          Thomas Zimmerer\inst{3}
          }

   \institute{
        {European Southern Observatory, Karl-Schwarzschild-Str.\ 2, Garching 85748, Germany}
         \and
        {Kavli Institute for Cosmological Physics, University of Chicago, Chicago, IL, 60637, USA}
         \and
        {OHB Digital Connect, Weberstra\ss e 21, D-55130 Mainz, Germany}
         \and
        {Independent Consultant, Kirchgasse 4, D-61184 Karben, Germany}
         \and
        {Institute of Theoretical Astrophysics, University of Oslo, P.O. Box 1029, Blindern, 0315 Oslo, Norway}
        \and
        {Instituto de Astrofísica and Centro de Astro-Ingeniería, Facultad de Física, Pontificia Universidad Católica de Chile, Santiago, Chile}
         \and
        {UK Astronomy Technology Centre, Royal Observatory Edinburgh, Blackford Hill, Edinburgh EH9 3HJ, UK}
         \and
        {Department of Physics and Astronomy, University of Pennsylvania, 209 South 33rd Street, Philadelphia, PA, 19104, USA}
         \and
        {Department of Technology Systems, University of Oslo, Gunnar Randars Vei 19, 2007 Kjeller, Norway}
         \and
        {Department of Physics, Cornell University, Ithaca, NY 14853, USA}
         \and
        {Department of Astronomy, Cornell University, Ithaca, NY 14853, USA}
         \and
        {INAF - Osservatorio Astronomico di Cagliari, 09047 Selargius, Italy}
         \and
        {European Southern Observatory, Alonso de Cordova 3107, Vitacura, Santiago, Chile}
         \and 
        {Instituto de Astrof\'{i}sica de Canarias (IAC), E-38205 La Laguna, Tenerife, Spain}
         \and
        {Universidad de La Laguna, Dpto. Astrof\'{i}sica, E-38206 La Laguna, Tenerife, Spain}
         \and
        {Departamento de F{\'i}sica de la Tierra y Astrof{\'i}sica e Instituto de F{\'i}sica de Part{\'i}culas y del Cosmos (IPARCOS). Universidad Complutense de Madrid, Av.\ Complutense, s/n, Ciudad Universitaria, 28040 Madrid, Spain}
      }

   \date{Received February 29, 2024; accepted January 14, 2025}

 
  \abstract  
   {The (sub)millimeter sky contains a vast wealth of information that is both complementary and inaccessible to other wavelengths.  Over half the light we receive is observable at millimeter and submillimeter wavelengths, yet we have mapped only a small portion of the sky at sufficient spatial resolution and sensitivity to detect and resolve distant galaxies or star-forming cores within their large-scale environments. For decades, the astronomical community has highlighted the need for a large, high-throughput (sub)millimeter ($\lambda\sim0.35-10$ mm) single dish. The Atacama Large Aperture Submillimeter Telescope (AtLAST), with its 50-m aperture and $2^\circ$ maximal field of view, aims to be such a facility. We present here the {preliminary} design concept for AtLAST, developed through {an EU Horizon 2020-funded design study}. Our design approach begins with a long lineage of (sub)millimeter telescopes, relies on calculations and simulations to realize the optics, and uses finite element analysis to optimize the {conceptual designs for the} mechanical structure and subsystems.  The demanding technical requirements for AtLAST, set by transformative science goals, {have motivated the design effort to} combine novel concepts with lessons learned from previous efforts.  The result is an innovative rocking chair design with six instrument bays, two of which are mounted on Nasmyth platforms, inside a large receiver cabin. Ultimately, AtLAST aims to achieve a surface accuracy of a $\leq 20~\mu$m root mean square half wavefront error, corresponding to the goal of a Ruze efficiency of $>50\%$ at 950~GHz. We conclude that a closed-loop metrology of the active primary surface will be required to achieve our surface accuracy goal. In the next phase of the project, we shall prototype and test such a metrology on existing platforms, with the goal of delivering a mature, construction-ready design by the end of this decade.}


   \maketitle
%
\section{Introduction}\label{sec:intro}

The desire to understand the Universe and how we came to be -- our {cosmic origins} -- is one of the most fundamental pursuits in astronomy.\footnote{We note that the use of the phrase {``cosmic origins''} does not imply any association, formal or informal, with any other effort.  While recent usage has included NASA's Cosmic Origins Program (officially founded in 2011; see \href{https://cor.gsfc.nasa.gov/}{https://cor.gsfc.nasa.gov/}) the film {\it ALMA---In Search of Our Cosmic Origins} (released in 2013; see \href{https://www.eso.org/public/videos/eso1312a/}{https://www.eso.org/public/videos/eso1312a/}),  and the Simons Observatory project motto ``Searching for our Cosmic Origins,'' the term has long been in use by the astronomical community. For instance, it appears in the book title {\it Our cosmic origins: from the Big Bang to the emergence of life and intelligence} \citep{Delsemme1998}, in the name of the Hubble Space Telescope's Cosmic Origins Spectrograph \citep{Bangert1997}, and in papers dating back to a century ago \citep[e.g.][]{1923ApJ....58..307B, 1925Natur.116..823M}.}
In order to address this fundamental question, we need a more complete multiwavelength view of our sky, including the ability to map the Universe at millimeter and submillimeter (hereafter (sub)millimeter) wavelengths, where approximately half of the light in the Universe as seen from our rest frame is detectable \citep[e.g.][]{Devlin2009, Vallini2023}.
Our ambitious concept for a new large aperture (50-meter), wide field of view (FoV; 2$^\circ$) single dish facility spanning (sub)millimeter wavelengths is motivated by ambitious observational capabilities that will allow us: (i) to perform the deepest, widest (100-1000 deg$^2$), and most complete imaging and spectroscopic surveys of the Galactic and extragalactic sky at a few arcsecond resolution, beating the confusion limits of previous experiments and resolving the cosmic infrared background (CIB); (ii) to study the morphology, kinematics, and chemistry of the multiphase gas -- and to study diffuse and low surface brightness gaseous structures -- in the interstellar and circumgalactic media of our own and other galaxies, which can extend across angular scales of degrees and carry the imprints of the baryon cycle, by observing them through multiple molecular and atomic emission lines; and (iii) to measure the thermodynamics and kinematics of the hot ionized gas in massive cosmic structures through the most sensitive subarcminute-resolution observations of the Sunyaev-Zeldovich (SZ) effect yet. The high surface brightness sensitivity, angular resolution, mapping speed, and imaging dynamic range of the Atacama Large Aperture Submillimeter Telescope (AtLAST)\footnote{\href{https://atlast-telescope.org/}{https://atlast-telescope.org/}} concept result directly from the needs of delivering on these science cases. 
None of them can be achieved through current or funded future facilities. Furthermore, they all require a high and dry site with excellent atmospheric transmission across the (sub)millimeter bands and a telescope able to make the most of such conditions. 

A brief introduction to AtLAST's scientific goals was presented in \cite{Klaassen2020}, who summarized the motivations and goals from the first AtLAST design study proposal submitted to the European Commission in 2019. These science goals were further broadened and updated in \cite{Ramasawmy2022} after consultation with a wider user community during the first years of the study, which commenced in March 2021. The recently delivered AtLAST Science Overview Report \citep{Booth2024b} as well as \cite{Booth2024a} summarize the key science goals at the end of this first phase.  These science cases are examined in greater detail through a large, concerted collection of dedicated papers (\citealt{Cordiner2024, DiMascolo2024, Klaassen2024, Lee2024, Liu2024, Orlowski-Scherer2024, vanKampen2024, Wedemeyer2024}).\footnote{See \href{https://open-research-europe.ec.europa.eu/collections/atlast}{https://open-research-europe.ec.europa.eu/collections/atlast}.}
Importantly, a complementary work by \cite{Akiyama2023} examines the crucial contribution AtLAST could make to very long baseline interferometry and the Event Horizon Telescope \citep{2019ApJ...875L...1E, 2022ApJ...930L..12A}.
As a side note, the science case presented in \citealt{Wedemeyer2024} motivates solar observations, which have been a key stretch goal for AtLAST since its inception; they now, however, constitute a compelling science driver, which translates into an observational requirement that impacts a few of the design choices discussed here. As a benefit, the same ability to safely point the telescope toward the Sun will also improve the overall observing efficiency by obviating the need for Sun avoidance.

AtLAST is conceived as a facility observatory that will serve a wide community of users for a long project lifetime ($>30$ years). It will be sited on the Llano de Chajnantor (Chajnantor plateau), approximately 5100~m above sea level in the Atacama Desert in northern Chile, a long-established site for (sub)millimeter observations \citep[see e.g.][]{Radford1998}.
AtLAST's diverse science goals drive the need for a large receiver cabin able to host multiple instruments that can access its large FoV and that can be upgraded throughout the lifetime of the observatory. In order to make the best use of sudden changes in weather conditions on the Llano de Chajnantor, the telescope operators need to be able to switch quickly (within a few minutes) between different instruments. This procedure of course needs to be carried out safely during both night and daytime operations. 

The AtLAST design concept builds upon nearly 50 years of experience with submillimeter observations and observatories, from pioneering projects like the Caltech Submillimeter Observatory \citep[CSO;][]{Leighton1977,Phillips1988}, the Swedish-ESO Submillimeter Telescope \citep[SEST;][]{Booth1987,Booth1989}, and the James Clerk Maxwell Telescope \citep[JCMT;][]{Hills1988, Hills1990}, to the Atacama Large Millimeter/Submillimeter Array \citep[ALMA;][]{Hills2008, Wootten2009}. AtLAST will be an excellent and much-needed complement for ALMA.  Along with upgrades to ALMA itself \citep[][]{Carpenter19,Carpenter2023}, AtLAST has the potential to contribute to keeping ALMA relevant for the foreseeable future: it will provide new targets and positions for high-resolution follow-up campaigns and offer the short {\it uv} baseline coverage needed for single-dish and interferometric data combination at high frequencies (e.g., \citealt{Plunkett2023, Bonanomi2024}). 

The concept of a large and truly (sub)millimeter single dish itself -- reaching frequencies of $\nu=300-950$~GHz -- has a long history dating back a number of decades \citep[e.g.,][]{Herter2004, Giovanelli2006, Woody2012, Kawabe2016, LSD_memo, Lou2020}, and has evolved and matured over this time.  A key step in this process is that we are now poised to deliver a design with truly transformative qualities, from the large FoV and optical throughput {(\'etendue)} that will ultimately allow megapixel-scale submillimeter cameras, to the power generation and energy storage and recovery systems necessary to carry out the project sustainably, to the metrology systems required to keep the beam performance exceptionally stable in a system not sheltered by a dome.

More globally, climate-change concerns and fuel price vulnerability faced by modern-day society have informed the development of the AtLAST concept, which has incorporated from the start a dedicated environmental sustainability study \citep[as is noted in, e.g.,][]{Klaassen2020}. The AtLAST project has driven new research into environmentally sustainable off-grid energy generation systems.  AtLAST's energy storage system needs to accommodate the high power demand of the observatory, providing a constant, reliable supply in all weather conditions, day and night \citep{Viole2023, Viole2024b, Viole2024}. This work may already be motivating other astronomical research infrastructures to either emulate these solutions, or produce similar alternatives, and so we hope that AtLAST's energy initiative efforts will have a wide impact on astronomy and throughout society as a whole. 
Building on combined approaches relying on previous studies from energy communities in the European Union (e.g., the Renaissance Project\footnote{\href{https://www.renaissance-h2020.eu/}{https://www.renaissance-h2020.eu/}.}), the AtLAST project is also considering the objectives of the local community and local stakeholders in the San Pedro de Atacama region in the design of its renewable energy system, contributing to the just and equitable use of energy resources in the area \citep{Valenzuela-Venegas2023}.

In this work, we present for the first time {a complete, preliminary} conceptual antenna design for AtLAST, developed during the first three years of the EU-funded design study (see Acknowledgments for further details), expanding on the overview provided by \cite{mroczkowski2023}. 
We emphasize that this conceptual design is simply the first phase, and we anticipate more development effort in the next phase to bring the project to a construction-ready status later this decade.
This paper is organized as follows. In Sect.~\ref{sec:designstudy}, we describe the overall motivation and scope of the AtLAST design study and discuss the key design requirements.  
In Sect.~\ref{sec:optics}, we describe the final optical design and how it was optimized.  
In Sect.~\ref{sec:antstruct}, we present the antenna structure and key results of the finite element modeling and analysis.
In Sect.~\ref{sec:achieving_perf}, we discuss and show how the optical performance, and the pointing and surface accuracy in particular, are achieved.
In Sect.~\ref{sec:insts}, we present the features of the receiver cabin and concepts for instrument installation and access.
And finally, in Sect.~\ref{sec:conclusions}, we provide our conclusions and discuss the next steps for the telescope design and the AtLAST project more generally.
As a convenience to the reader, we list the acronyms and abbreviations we use in Table \ref{tab:jargon}.

\section{The AtLAST design study}\label{sec:designstudy}

The key design goals of AtLAST can be summarized as follows: we aim to build a facility with a large collecting area, fast mapping speeds supporting flexible mapping strategies, high angular resolution, and the ability to observe up to frequencies of $\approx 950$~GHz.  Furthermore, in order to serve a multitude of science goals, this facility must feature a large receiver cabin capable of housing multiple massive instruments. The space allocated for each receiver is large, as two of them must ultimately be able to fill a FoV roughly two orders of magnitude larger than any previous large (sub)millimeter single dish has had, while the mounting points for the smaller receivers must also be capable of hosting instruments that fill a significant portion of the FoV.

While noteworthy pathfinding studies have been undertaken \citep[e.g.,][]{Woody2012, Kawabe2016}, {the AtLAST design is uniquely optimized for sustainability and upgradability.}  We note that AtLAST is in part inspired by such pathfinding studies as the original CCAT design, which undertook a longer and more extensive design effort than the one we present here.\footnote{For reference, the original CCAT design featured an enclosed 25-meter telescope reaching similar frequencies as AtLAST and had a limiting FoV $\approx 1^\circ$ at 850~GHz \citep[e.g.,][]{Woody2012}, while the Large Submillimeter Telescope \citep[LST;][]{Kawabe2016} featured a domeless 50-meter telescope with a limiting FoV $\approx0.7^\circ$ at the nominal upper frequency of 420~GHz for the full collecting area. 
We also note some overlap with the contemporary concept for a 60-meter submillimeter telescope reported in \cite{Lou2020}, which aims for an upper frequency of 500~GHz and a 1$^\circ$ diameter FoV.} 
To this end, a project including the telescope design -- structural engineering, optical design, finite element analysis (FEA), and end-to-end modeling -- was proposed to the European Union’s Horizon 2020 research and innovation program, and in March 2020 it was awarded.
While the project officially commenced a year later, in March 2021, the telescope design work began in earnest in July 2020 with the optical design considerations later summarized in \cite{AtLAST_memo_1}.
{We note for completeness that the 25-meter CCAT design effort resulted in a comparatively more detailed dynamical analysis of the antenna structure \citep{Woody2012} than currently available for AtLAST. As the AtLAST design progresses throughout the next phase of the project, we aim to deliver such an analysis.}

\subsection{Design approach}\label{sec:approach}

As is described in Sect.~\ref{sec:optics} and further detailed in \cite{Gallardo2024}, the AtLAST optical design goal was to maximize the FoV while minimizing the mass (for a fixed 50~m diameter primary mirror). This in turn drives many of the structural choices for the antenna concept.

Our overall design approach was iterative, since many of the design choices could not be treated independently.
Our structural design approach began with the optical layout and, building upon industrial experience with previous telescope designs, followed the principles of homologous support  \citep[described in][]{BaarsKaercher2018} to find a working solution for the primary mirror support structure. Ultimately this converged on a concept for the overall antenna structure.  From there, we produced a model for the telescope (Sect.~\ref{sec:CAD}), which was then iterated through finite element modeling (Sect.~\ref{sec:fem}) until {it converged} on a working structural design for the telescope.  Further tests and modeling were then used to assess the telescope design properties and, if necessary, further adjust and refine the structural design.

\subsection{Design requirements}\label{sec:requirements}

The key science drivers for AtLAST \citep{Booth2024b, Booth2024a} lead to a set of demanding observational capabilities, which in turn set most of our key design requirements; a number of other requirements are also set by the demanding environment or the desire to facilitate maintenance and operations.
The overall design goals and requirements for AtLAST were presented in \cite{Klaassen2020}.  
We provide an updated version of the key technical requirements and design solutions for AtLAST in Table~\ref{tab:design} in order to guide the general discussion in this work. 
Of the requirements in Table~\ref{tab:design}, the following subsections discuss those most crucially driving the telescope design.

\begin{table}[tbh]
\centering                          
\caption{Key technical requirements and design solutions for AtLAST}
\begin{tabular}{l | c}        
\hline\hline                 
Parameter & Value \\    
\hline                       
Wavelength ($\lambda$) range                & 0.3-10~mm      \\
Primary mirror diameter                 & 50~m \\
Field of view (FoV)                             & 2$^\circ$ (1$^\circ$)\tablefootmark{$\dag$}\\
Number of instruments                   & $\geq 5$             \\               
Effective focal length                      & $\approx 100$  m    \\
Number of mirrors                       & $\leq 4$             \\               
Number of segments                          & $\approx 400$        \\
Sizes of segments                               & $\approx 5$   m$^2$ \\
Total collecting area                       & $\approx 2000$  m$^2$\\
Optical surface accuracy                & 20-30~$\mu$m         \\
Surface coating                                 & similar to ALMA      \\
Optical design                          & Cassegrain-Nasmyth   \\
Description of active optics                & active surface + metrology \\ 
Actuator precision                              & $\approx 10~\mu$m    \\
Mechanical pointing accuracy\tablefootmark{$\star$}       & 2\farcs5        \\
Scan speed                              & 3$^\circ \rm s^{-1}$ \\
Acceleration                            & 1$^\circ \rm s^{-2}$ \\
Elevation (EL) range                    & $20 ^\circ - 90^\circ$ \\
Azimuthal (AZ) range                    & $\pm 270^\circ$ \\
Mount type                              & AZ-EL       \\                
Support structure material              & steel \& invar       \\               
\hline 
\end{tabular}
\tablefoot{
\tablefoottext{$\dag$}{A diameter of 1$^\circ$ is the minimum requirement for the FoV at 850~$\mu$m, while at longer wavelengths ($\geq2$~mm) the FoV diameter requirement is 2$^\circ$.}
\tablefoottext{$\star$}{Here we refer to the ability to point at and track a given astronomical position on the sky after slewing from another arbitrary position elsewhere on the sky.  The pointing will be referenced to celestial coordinates, but this telescope pointing requirement does not, for instance, include the potential gains from the use of a sophisticated all-sky astronomical pointing model like those in use on other platforms. See discussions in Sects.~\ref{sec:pointing_req} and \ref{sec:pointing_accuracy}.}
}
\label{tab:design} 
\end{table}

\subsubsection{Wavelength range and surface accuracy}\label{sec:surface_acc_req}

AtLAST aims to cover approximately the same wavelength range ($\lambda  \approx 0.3-10$~mm, $\nu \approx 30-950$~GHz) as ALMA, with the lower end of the frequency range extended down to 30~GHz  as requested by some of the key science goals, such as those relying on AtLAST's complementarity with current and planned Cosmic Microwave Background (CMB) survey experiments (albeit at $> 8.33\times$ finer angular resolution and $>69.4\times$ point-source sensitivity).  

The high frequency goal for AtLAST drives not only the site selection, but also the surface accuracy {or, more precisely, the root mean square (RMS) half wavefront error (HWFE),} which limits the aperture efficiency at the highest attainable frequency. 
Here, an aperture efficiency of $\geq$50\% that of the lowest frequencies is generally considered acceptable. 
As a first approximation in this work, we use the \cite{Ruze1966} formula to estimate what surface accuracy is required to achieve a sufficiently high aperture efficiency. 
We note that our surface accuracy goal of 20~$\mu$m {RMS HWFE} (Table~\ref{tab:design}) is comparable to that achieved by the 12-meter ALMA antennas. This level of {surface} accuracy has been shown to be sufficient for observations up to $\approx 950$~GHz, making use of the highest frequency atmospheric windows available from Llano de Chajnantor.
The surface accuracy goal considers the errors caused by surface deformations on small scales (2-5~m), while we note that the large scale deformations ($>5$~m) of the primary reflector will be compensated by an active surface working in a closed-loop mode with a sensor system as well as by the pointing model.
On the Chajnantor plateau, the best observing conditions are typically available at night and early mornings \citep[e.g.,][]{Cortes2020, GomezToribio2021, Morris2022, Morris2024}, and so we refer to 20~$\mu$m as our nighttime surface accuracy goal.
We also set a strict requirement of 30~$\mu$m {RMS HWFE} for daytime observations, when thermal effects are worse and both wind speed and precipitable water vapor (PWV) are generally higher. 

\begin{figure}[tbh]
        \centering
        \includegraphics[width =\hsize]{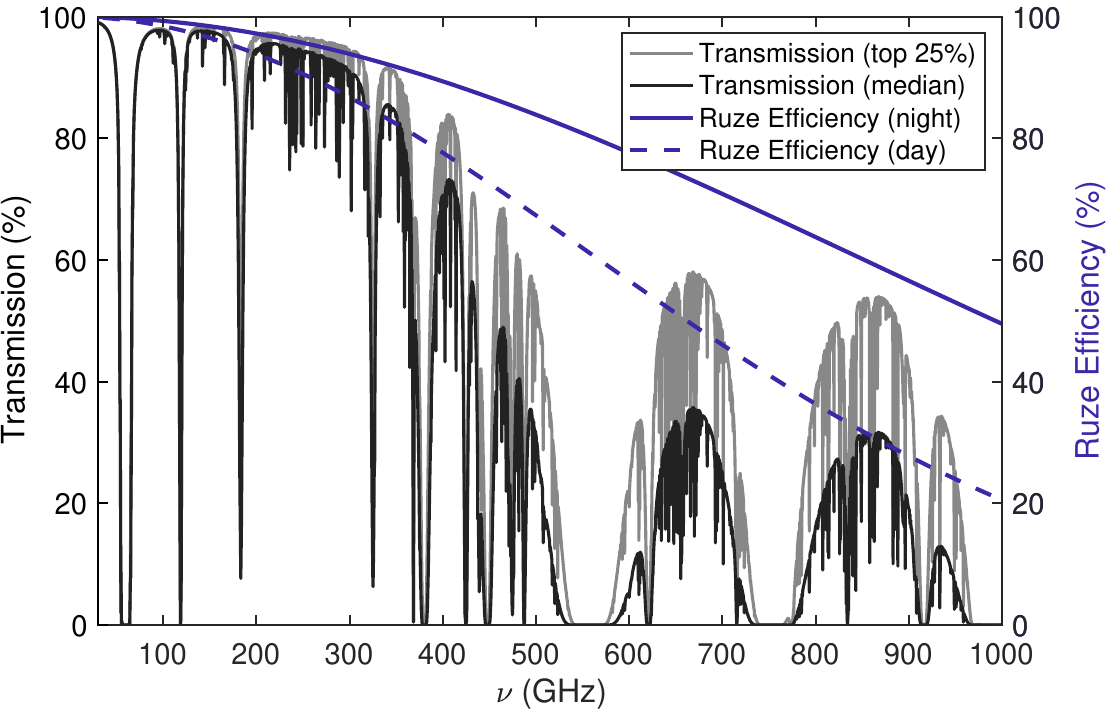}
        \caption{Annual average transmission of the atmosphere in AtLAST's bands using the median and top quartile conditions for the ALMA site. Plotted on the right axis is the Ruze efficiency for the 20~$\mu$m nighttime surface accuracy goal, and that for the 30~$\mu$m daytime surface accuracy expectation.}
        \label{fig:Ruze}
\end{figure}
    
In Fig.~\ref{fig:Ruze}, we show the results of our Ruze efficiency calculation for both the 20~$\mu$m nighttime surface accuracy goal and for the 30~$\mu$m daytime surface accuracy expectation. 
We compare this to the atmospheric transmission, computed using the \texttt{am} code \citep{am:2018}. 
{Here we note that telescope surface accuracy will not be the sole limiting factor for high frequency observations, as the losses in efficiency due to Ruze scattering can be expected to be comparable to the losses due to atmospheric transmission.}
\cite{Gallardo2024} and \cite{Puddu2024} provide more details about the optics, including calculations of the overall optical performance as well as a full physical optics calculation of the beam and cross-polarization performance. 
The nuances and limitations of applying the Ruze formula are discussed further in Sect.~\ref{sec:achieving_perf_general}.

\subsubsection{Primary mirror diameter}

{The choice of aperture size for the AtLAST primary mirror is driven by the key science goals mentioned Sect.~\ref{sec:intro}, which require both angular resolution and sensitivity (i.e.,\ collecting area).  However, the aperture size must be balanced against the introduction of unnecessary risks to the project itself.}
While larger aperture sizes may in principle be achievable \citep[see e.g.,][]{KaercherBaars2014}, this choice represents a trade-off between the minimum diameter necessary to deliver AtLAST's key observational capabilities {versus the maximum size demonstrated for an unenclosed structure operating at nearly comparable wavelengths.  
In particular, the Large Millimeter Telescope/Gran Telescopio Millimetrico {Alfonso Serrano} \citep[LMT/GTM;][]{KaercherBaars2000} is a 50-meter telescope with a maximum frequency of up to $\nu \sim 350$~GHz, which is only a factor $\lesssim 3$ from AtLAST's high frequency requirement.}

When considering the science cases for AtLAST, we found 50 meters to be the minimum size to fully resolve the Poissonian CIB at our highest frequencies ($\sim 950$~GHz), allowing the emission from sources to be distinguished from each other instead of being fundamentally limited by confusion \citep[see e.g.,][]{vanKampen2024}. 
{Similar improvements regarding the ability to distinguish multiple sources and features, as well as to separate their signals from those of their surrounding environments, can also be expected for all of AtLAST's observational goals, Galactic and extragalactic.}
A more stringent lower limit on the aperture size is also set by the desire to provide sufficient overlap in Fourier space with ALMA's main 12-meter Array\footnote{ALMA comprises three arrays \citep[e.g.,][]{Hills2008, Wootten2009}.  Here, we focus on the main, fifty element interferometric array.} for data combination that results in  high spatial fidelity imaging at even smaller scales than those that will be probed by AtLAST alone.  As is shown in, for example, \cite{Frayer2017} and \cite{Plunkett2023}, the minimum requirement for reliable data combination is that the single dish be $\geq 3\times$ the shortest projected baseline length in the array.  For the main 12-meter Array of ALMA in a compact configuration (i.e.,\ observing to the antenna shadowing limit), this diameter would be 36 meters. 
Regardless, our 50-m goal ensures that we are able to achieve more than sufficient overlap in Fourier space with both ALMA and ngVLA as well as any high-frequency upgrades to the Square Kilometer Array (SKA).\footnote{See for example \href{https://www.skao.int/sites/default/files/documents/d38-ScienceCase_band6_Feb2020.pdf}{SKA Memo 20-01}.}

\subsubsection{Pointing accuracy}\label{sec:pointing_req}
 
The mechanical pointing accuracy is the residual deviation of the telescope's orientation relative to an astronomical source when slewing from an arbitrary location elsewhere on the celestial hemisphere. The pointing stability is the accuracy at which the pointing direction to the source can be maintained over the maximum time period of a continuous measurement of the source. The demands on the pointing stability are usually higher than the mechanical pointing accuracy during the acquisition of the source.
Due to the large surface area of the primary mirror, the requirement of high frequency operations, and the unsheltered structure of AtLAST, the pointing requirements may be some of the most challenging requirements for AtLAST.

{Like the large aperture mm-wave telescopes mentioned below,} we anticipate that AtLAST will rely on an astronomical pointing model for corrections to achieve its pointing accuracy requirements, ensuring a source is well within the width of the main beam of the (sub)millimeter telescope. 
{We therefore set a corrected pointing requirement of 0.5-1\arcsec\ at the highest frequencies for AtLAST. Prior experience shows that the mechanical pointing error can be improved upon by a factor $\sim 7\times$ using an astronomical-source-based pointing correction/look-up table. We degrade this improvement factor to 2.5-5$\times$ to be conservative, yielding a 2\farcs5 mechanical pointing requirement. \cite{Reichert2024} discusses in detail the 100\arcsec\ mechanical pointing accuracy passively achievable and lists approaches that may enable this 2\farcs5 requirement to be met.}

{Our expectation of improving by a factor of $\sim$ 2.5-5$\times$ our mechanical pointing accuracy requirement by introducing a pointing look-up table is based on demonstrated results from existing large (sub)millimeter telescope facilities successfully applying this strategy.} For instance, the 64-m Sardinia Radio Telescope \citep[SRT;][]{Prandoni2017}, which will observe at frequencies up to 116~GHz, achieves 13\arcsec\ mechanical  pointing accuracy, but has demonstrated 2\arcsec\ overall pointing accuracy when corrected with a standard astronomical pointing model.  
Similarly, the 100-m Green Bank Telescope (GBT), which observes up to 116~GHz with a 8\farcs5 beam, was designed to achieve an uncorrected pointing accuracy $\approx 9\arcsec$.
\cite{White2022} demonstrated, however, that they could correct the root mean square error to 1\farcs2 accuracy using an astrometrically corrected pointing model for the GBT.  
This is a dramatic improvement over the previous pointing-model-corrected value of 2\farcs8 \citep{Prestage2009}. 
These commendable levels of improvement (by factors of $\approx 7\times$ better) indicate that our expected improvement relative to mechanical pointing for AtLAST should also be achievable.

Finally, we note that we anticipate that AtLAST will use a static pointing error correction model combined with closed-loop control compensation techniques (e.g., relying on the concept of flexible body control as detailed in \citealt{BaarsKaercher2018}). 
We describe our approach to achieving the mechanical pointing accuracy requirement in Sect.~\ref{sec:pointing_accuracy}.

\subsubsection{Field of view}\label{sec:FoV}

{AtLAST's large FoV, which---together with collecting area---will} enable fast mapping speeds \citep[][]{Cicone2019,Klaassen2020}, is one of the key drivers for the optical design (Sect.~\ref{sec:optics}), as requested by the majority of the science cases \citep{Booth2024b,Booth2024a}.
The large FoV places significant demands on the overall antenna structure (Sect.~\ref{sec:antstruct}), ultimately driving us to the choice of a rocking chair design for the elevation structure (described in Sect.~\ref{sec:design_approach}).

The FoV of an imaging (sub)millimeter single dish telescope like AtLAST determines two key parameters: the mapping speed (as is noted above, and discussed here), which scales as the number of beams on sky, and the largest angular scale that can be recovered in the observation after removal of the atmospheric signal \citep[see][]{vanMarrewijk2024b}.
Following \cite{Cordes2008} and \cite{Wilson2013}, we define a figure of merit (FoM) for survey mapping speed as
\begin{equation}
    \textrm{FoM} = \eta \, N_\textrm{FoV} \, \Omega_\textrm{FoV} \, \Delta \nu \left( \frac{A_\textrm{eff}}{2 k_B T_\textrm{sys}} \right) ^2.
\end{equation}
Here, $\eta$ is an efficiency factor which can be assumed to be approximately the same for telescopes with similar surface accuracy, $N_\textrm{FoV}$ is the number of FoVs probed simultaneously, $\Omega_\textrm{FoV}$ is the solid angle of the FoV, $\Delta \nu$ is the bandwidth, $A_\textrm{eff}$ is the effective area, $k_B$ is the Boltzmann constant, and  $T_\textrm{sys}$ is the system noise temperature.  
For both AtLAST and ALMA, we set $N_\textrm{FoV} = 1$; ALMA is single beam, and has not been upgraded to host focal plane arrays, while we consider AtLAST to have a single, large FoV.
The ratio of the collecting area of AtLAST to that of ALMA is 0.34 assuming 50 ALMA 12-m antennas.\footnote{We note that the current minimum number of antennas for ALMA 12-meter Array observations is 43 \citep{almatechnicalhandbook}.}
Therefore, assuming the same bandwidth, efficiency, and $T_\textrm{sys}$, the advantage for AtLAST lies in $\Omega_\textrm{FoV}$. For instance, at 850~$\mu$m we expect the recoverable\footnote{The recoverable FoV is greater than or equal to the diffraction limited FoV of the telescope itself, and depends on optical corrections in the instrument.} FoV of AtLAST to be $\approx 80^\prime$ in diameter (see \citealt{Gallardo2024}), yielding $\Omega_\textrm{FoV, AtLAST}/\Omega_\textrm{FoV, ALMA} \gtrsim 90000$ and $\rm FoM_\textrm{AtLAST}/FoM_{ALMA} \gtrsim 10^4$.
While we can expect an aperture efficiency $A_\textrm{eff}$ comparable to ALMA (see Sect.~\ref{sec:surface_acc_req}), further improvements to this figure of merit can be realized through, for example, wider receiver bandwidth $\Delta \nu$, instruments that allow simultaneous observations in multiple bands, or by achieving lower receiver noise temperatures, which in turn would result in lower $T_\textrm{sys}$ \citep[see][who note the receiver noise temperatures in all current ALMA bands are factors $\sim 4-10\times$ higher than the quantum noise limit]{Carpenter2023}.

\subsubsection{Scanning requirements}\label{sec:scanning}

{AtLAST's ability to scan and accelerate quickly will be crucial both for mapping large areas of the sky and recovering large angular scales \citep[see Sect.~\ref{sec:intro} and][]{Booth2024a, Klaassen2024, DiMascolo2024}.  Given the size of AtLAST, fast acceleration and velocity will likely be two of the most important structural engineering challenges for AtLAST.  At the present time, the full finite element modeling results for how the antenna structure will respond to speed, acceleration, and impulse are not available; these will be explored in the next phase of development.

As has been noted in the literature on (sub)millimeter mapping \citep[see e.g.,][]{Dunner2013, Morris2022, vanMarrewijk2024b}, the recovery of faint continuum emission as well as faint line emission whenever the line frequency is not known a priori depends critically on the ability to modulate the astronomical signal faster than the atmospheric signal changes.} 
At 12 meters in diameter (Table~\ref{tab:optical_parameters}), the secondary mirror of AtLAST is too large to serve as a wobbler (though see Sect.~\ref{sec:wobbler} for discussion of possible solutions for a smaller-FoV wobbler in the receiver cabin). 
This drives the requirement for AtLAST to scan quickly in order to achieve this modulation, while the operational requirement to minimize observational overheads drives the need for rapid acceleration. 

In practice, the timescale at which the atmosphere no longer dominates\footnote{We note that the noise power spectrum of a bolometer or heterodyne receiver generally also has a component of `pink' noise.  This noise, generally related to drifts in detector gain and thermal properties of the instrument, is higher at lower temporal frequencies (on longer timescales) than at higher ones (on shorter timescales).  In the case of phase-incoherent detectors -- such as bolometers or kinetic inductance detectors -- this is often referred to as $1/f$ noise \citep[e.g.,][]{Mather1982}; for phase-coherent heterodyne instruments (e.g., receivers relying on amplifiers), this is usually measured as the Allan variance \citep[e.g.,][]{Yagoubov2020}.
In practice, this also drives the requirement to modulate the signal quickly enough that the astronomical signal is out of the pink noise-dominated portion of the instrument's noise power spectrum, and generally in a Gaussian white noise (i.e.,\ featuring a more flat power spectrum) portion, where the amplitude of the total cumulative noise is lower \citep[see e.g.,][]{Dunner2013, vanMarrewijk2024b}.} the noise power spectrum of the data is of order $\sim 0.3$~seconds \citep[e.g.,][]{Dunner2013}, which implies a scan speed of $\gtrsim 1.7^\circ \rm s^{-1}$ would be required to recover the $\sim 0.5^\circ$ scales noted in several AtLAST science cases.
However, using data from the Atacama Cosmology Telescope (ACT) and sophisticated modeling of the atmospheric turbulence and time-evolution, \cite{Morris2022} showed that the frequency at which the atmospheric noise meets the detector white noise level can be modulated by scanning.
In particular, they found that the noise on larger scales was lower for 1.5$^\circ \rm s^{-1}$ scans following the wind direction than for the same speed of scans in the opposite direction. 
\cite{Morris2022} posited that the optimal scan speed would be that which keeps the atmospheric signal stationary in the time domain (i.e.,\ the ideal case in the so-called ``frozen sky'' approximation discussed in \citealt{Coerver2024}), facilitating for instance filtering and common mode subtraction.
More recently, \cite{Coerver2024} reported measurements in both intensity and polarized atmospheric noise for the South Pole Telescope, finding the noise was higher for scans taken at higher telescope elevation angles but constant azimuthal speeds. They attribute the higher noise levels to the declining angular speed on sky at higher elevation angles of the observations, meaning the frozen sky approximation was not satisfied in these cases. 

\begin{figure}[tbh]
    \centering
    \includegraphics[width=\hsize, trim={0.1cm 0.1cm 0.1cm 1.0cm},clip]{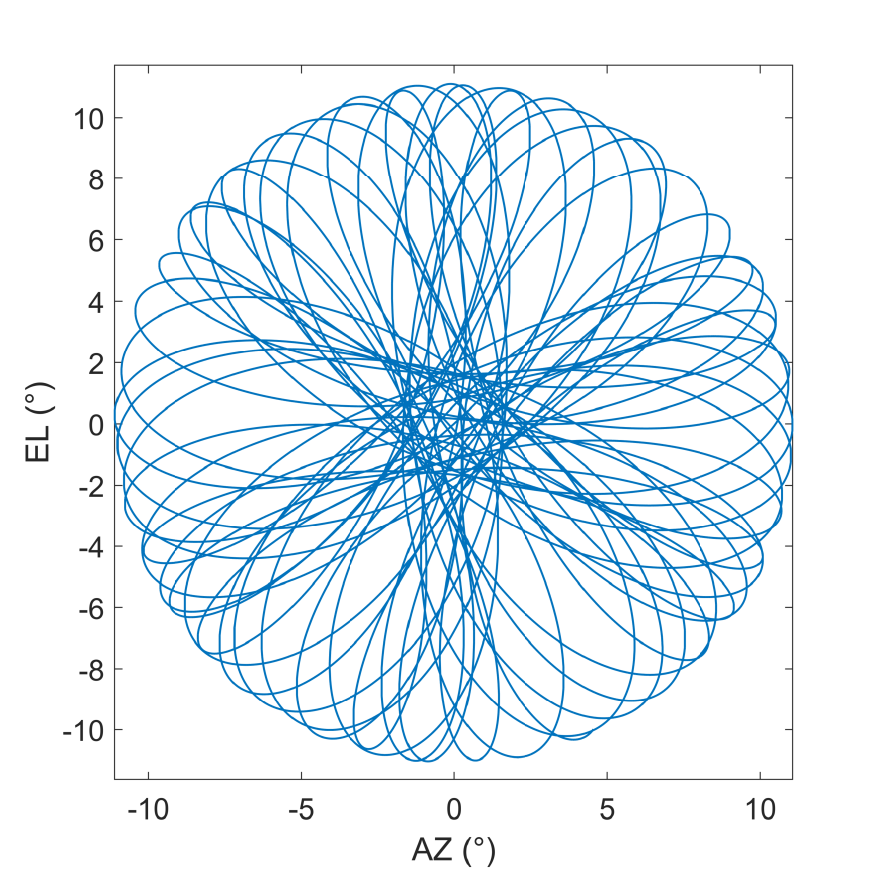}
    \includegraphics[width=\hsize, trim={0.1cm 0.5cm 0.1cm 0.5cm},clip]{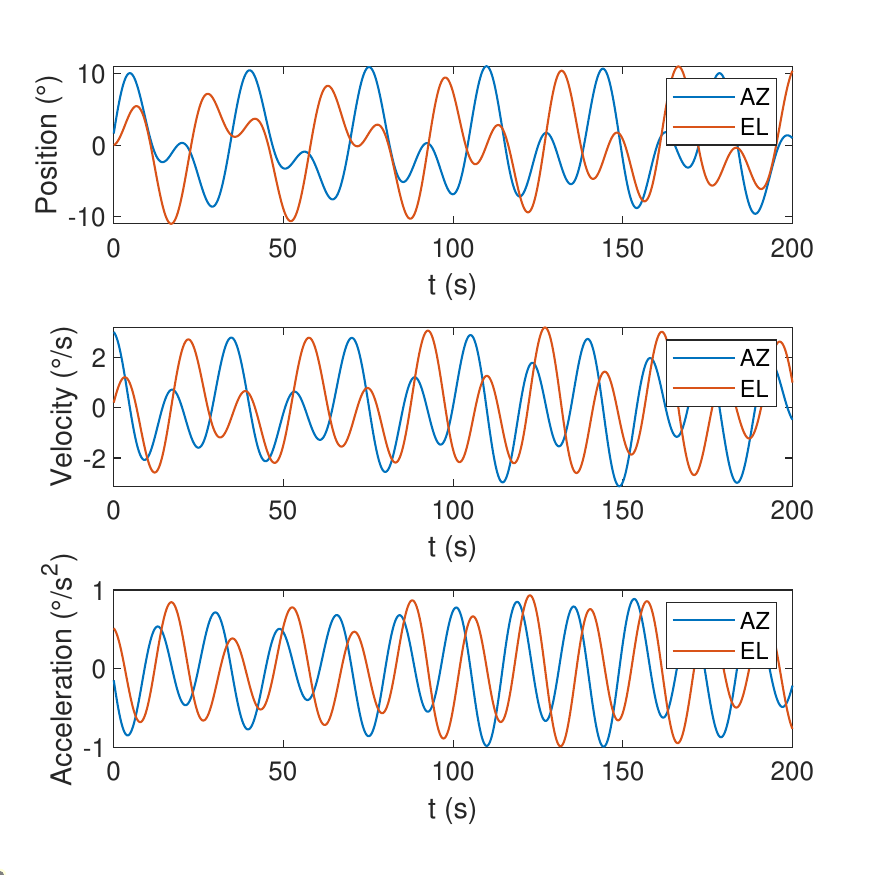}
    \caption{Representative Lissajous daisy scan for AtLAST that, at its maximum velocity and acceleration, reaches the kinematic constraint and antenna drive limits. The top panel shows the relative offsets in elevation and azimuth with respect to the field center, and the lower plots show the corresponding position, velocity, and acceleration as a function of time. Figure~\ref{fig:power_demand} shows the associated power demands of this example scanning pattern.}
    \label{fig:liss_scan}
\end{figure}

While more work to fully simulate these effects will be required \citep[e.g., through the tools and methodology described in][]{vanMarrewijk2024b}, we use the above results as heuristic guidelines when setting AtLAST scanning speed requirement and conservatively choose a maximum scan speed of 3$^\circ \rm s^{-1}$.  This includes a margin that may allow AtLAST to satisfy the frozen sky approximation in faster wind conditions as well as higher elevation angles.
As a precaution, the maximum allowed speed linearly declines from $3^\circ \rm s^{-1}$ to $0^\circ \rm s^{-1}$ in the last $5^\circ$ of the AZ/EL ranges (i.e., $\mathrm{EL} <25^\circ$, $\mathrm{EL} > 85^\circ$, and $|\mathrm{AZ}| > 265^\circ$).
To minimize observational overheads, we choose an acceleration limit of 1$^\circ \rm s^{-2}$, which places strong demands on the antenna's drives and stringent requirements on the telescope stiffness, in particular that of the secondary support structure during turnaround. We discuss the impact of accelerations on the structure in Sect.~\ref{sec:design_approach}, and we discuss the energy requirements and how to reduce the demand in Sect.~\ref{sec:energy} and \cite{Kiselev2024}.

While the motivations for fast scanning discussed above largely focus on survey modes using azimuthal scans at constant elevation \citep[e.g.,][]{Swetz2011}, AtLAST aims to be a user facility.  As such, AtLAST will also support a wide range of scanning patterns like those implemented in other telescopes, including simple daisy scans, classic Lissajous scans, Lissajous daisy scans \citep[e.g.,][]{Romero2020, vanMarrewijk2024b}, and several tracking modes, such as sidereal tracking or more rapid tracking of solar system objects like comets, asteroids, and near Earth objects.
A representative Lissajous daisy scanning pattern for AtLAST, which attains the maximum allowed acceleration and speed, is shown in Fig.~\ref{fig:liss_scan}. This case, which hits the peak acceleration values multiple times in a short period, is discussed further in Sect.~\ref{sec:energy}.

\subsubsection{Sustainability}\label{sec:sustainability}

While not a formal requirement of the telescope design itself, a crucial goal and driving philosophy in the AtLAST design study has been to maximize the environmental as well as social sustainability of the project. This is in line with the recommendation of, for example, the United Nations Intergovernmental Panel on Climate Change \citep[IPCC; see e.g.,][]{IPCC2023}, who recommend achieving carbon neutrality before 2050.
To this end, the AtLAST design study included a crucial work package on delivering solutions for sustainably powering the facility \citep{Viole2023, Viole2024}. The two prospective sites\footnote{See the \href{https://atlast-telescope.org/design-study/wp3-site/d3.1_site_selection_criteria.pdf}{AtLAST site selection criteria report}.} being considered for AtLAST, both of which at an altitude of 5050 meters above sea level in the Atacama Desert in Northern Chile, are not connected to the power grid. While a number of astronomical projects in the region are now implementing solar arrays that will vastly reduce their reliance on carbon-based fuels, AtLAST is the first, to our knowledge, designed to be fully powered by renewable energy sources from its inception.

AtLAST is also among the first astronomical design studies to publish a life cycle assessment of its possible energy system setups, which are described in Sect.~\ref{sec:energy}. \cite{Viole2024b, Viole2024} found that systems relying on large shares of solar photovoltaic generation to have significantly lower carbon footprints over the system's life cycle compared to today's diesel generators, while requiring more water and metal resource use. Further work is needed to also include the telescope itself within the scope of such an environmental assessment, as proposed by \cite{Knodlseder2022}. \cite{Valenzuela-Venegas2023} have also addressed social sustainability and energy justice through including the energy system preferences of stakeholders of the nearby community of San Pedro de Atacama in the design process.

In addition to the sustainability study mentioned above, the telescope itself will implement innovative concepts to reduce its power demand. As one of the design drivers is fast scanning and acceleration (Table~\ref{tab:design} and Sect.~\ref{sec:scanning}) of a massive structure, it follows that the energy required for such motions is also high. We therefore developed an energy recovery system concept based on supercapacitors, which is expected to allow regenerative braking and thus drastically improve the energy efficiency of the telescope's motion. 

We provide an overview of the power demands and the energy recovery system in Sect.~\ref{sec:energy}, while the full description can be found in \cite{Kiselev2024} and the concepts for power generation and energy storage were presented in \cite{Viole2023, Viole2024b, Viole2024}.

\section{AtLAST antenna optics}\label{sec:optics}

The optical design for AtLAST takes a Ritchey-Chr{\'e}tien hybrid Cassegrain/Nasmyth approach, with the primary and secondary mirrors (M1 and M2, respectively) located on axis, and a flat, folding tertiary mirror (M3)\footnote{Here we mean “folding” in an optical sense: it is flat and does not change the focal length, hence it “folds” the beam.  The tertiary mirror does not fold mechanically.} that allows fast instrument selection.
Like many of the current generation of (sub)millimeter telescopes that are smaller (generally $\sim 6$~meter) than AtLAST but achieve high throughputs (FoV times collecting area) and correspondingly high mapping speeds, AtLAST features fast optics, with a primary mirror focal ratio of $\approx 1/3$, and a focal ratio at the instrument of $\approx 2.6$. Here we refer to what is normally termed the `focal plane' as the `focal surface,' in recognition of its significant curvature.

A focal ratio of $\approx 2.6$ was chosen both to keep the physical scale of the focal surface as small as reasonably possible and to maximize compatibility with existing receiver designs from, for example, the Simons Observatory \citep[SO;][]{Zhu2021, Bhandarkar2022}, CCAT \citep{Vavagiakis2018}, and CMB-S4 \citep{Gallardo2022, Gallardo2024tma} with relatively few changes.
In the next sections, we describe the evolution of the design, followed by a description of the final optical design for AtLAST.

\subsection{Evolution of the optical design}\label{sec:optical_design_evolution}

We initially considered several approaches for the optical design before making a down-selection, as detailed in AtLAST Memo \#1 \citep{AtLAST_memo_1}. 
These included a Cassegrain, a Nasmyth, a three-mirror symmetric approach similar to that taken by the Vera Rubin Observatory, and a three-mirror off-axis approach inspired in part by the design of the 100-m Green Bank Telescope (GBT; \citealt{White2022}), which has two mirrors, and the CMB-S4 Three Mirror Anastigmatic design \citep[TMA; ][]{Padin2018, Gallardo2024tma}.

Due to the size, mass, and overall feasibility considerations outlined in the memo, we converged on a Ritchey-Chr{\'e}tien design which optimizes the compactness of the structure while achieving a wide FoV.
For comparison, AtLAST's geometric FoV is roughly $500\times$ larger than that of the 50-meter LMT/GTM.
We note that historically, though the Ritchey-Chr{\'e}tien design was developed in the early 1900's for visible wavelength telescopes, it was developed essentially for the same reason AtLAST chose to adopt it now. The Ritchey-Chr{\'e}tien optical design reduces coma and maximizes the FoV, allowing much larger images to be taken than with traditional parabolic reflectors \citep{Wilson1996}. In this sense, AtLAST is among the first large (sub)millimeter telescopes to be designed specifically for wide field imaging.

\subsection{The final optical design}\label{sec:final_optical_design}

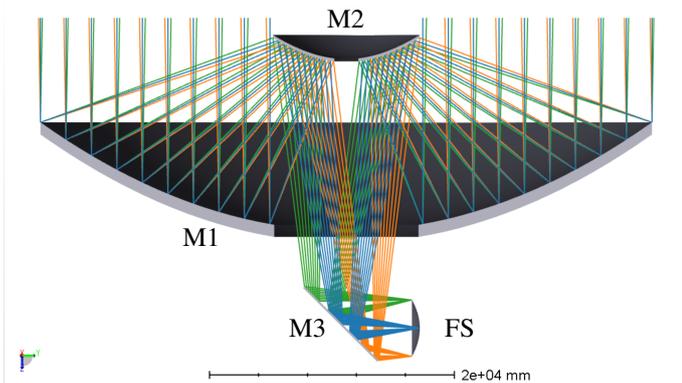
\begin{figure}[tbh]
        \centering
    \begin{subfigure}[b]{\hsize}
                \centering
            \begin{tikzpicture}
    \draw (0, 0) node[inner sep=0] {\includegraphics[width =\hsize, trim={0.1cm 0.1cm 0.1cm 1.5cm},clip]{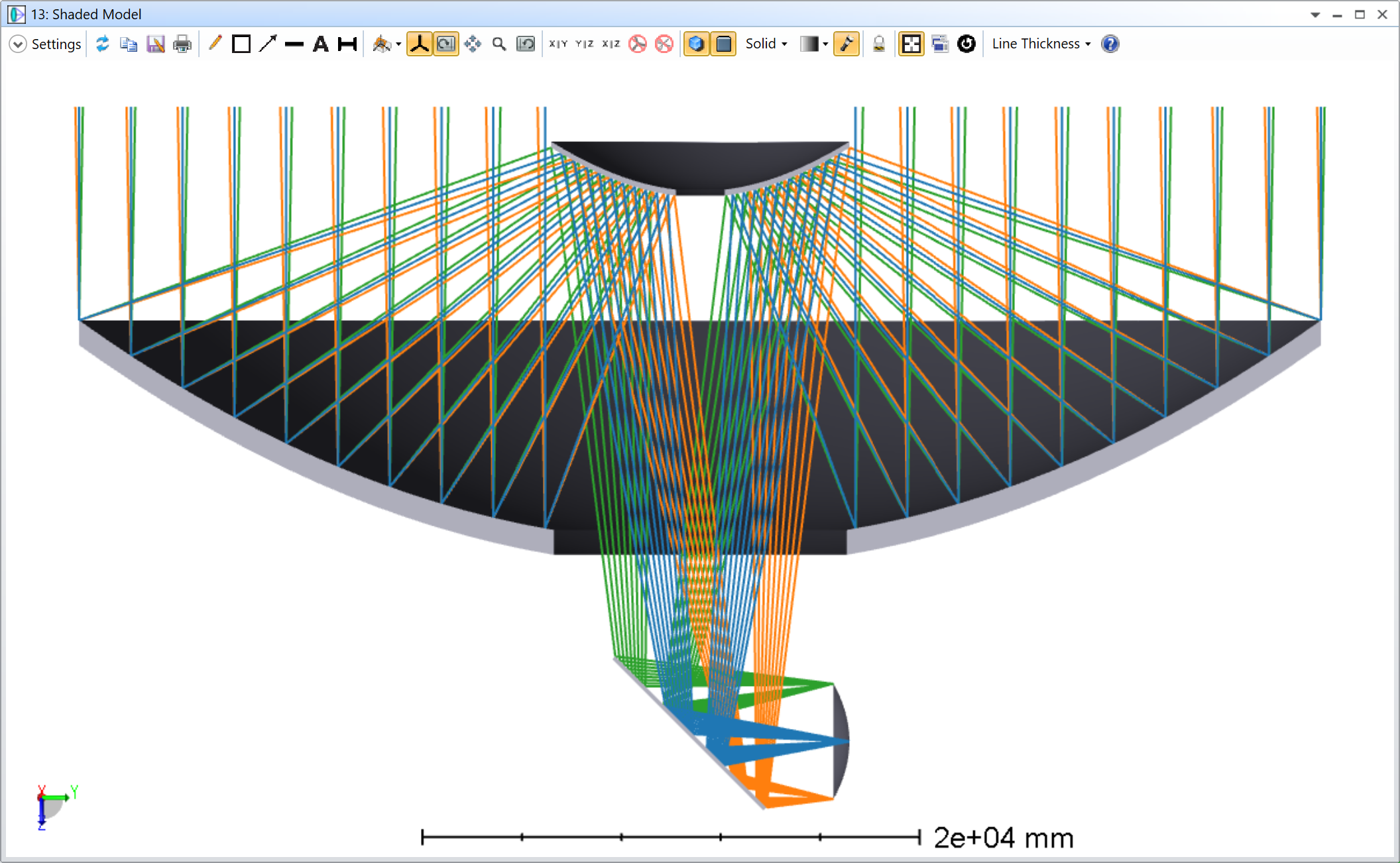}};
    \draw (-1.9,  -0.6) node {M1};
    \draw (0,  2.3) node {M2};
    \draw (-0.5, -1.8) node {M3};
    \draw (1.5, -1.8) node {FS};
\end{tikzpicture}
    \end{subfigure}  
        \caption{The optical layout for AtLAST. Mirrors M1, M2, and M3 focus light from the sky onto the focal surface FS.}
        \label{fig:atlast_optics}
\end{figure}

Fig.~\ref{fig:atlast_optics} presents AtLAST's optical layout, and the optical parameters are summarized in Table \ref{tab:optical_parameters}.  Here, the back focal distance is defined as the distance along the optical path from the central opening in the primary mirror surface to the center of the focal surface.

\begin{table}[tbh]
\caption{Optical parameters of AtLAST.}             
\centering                          
\begin{tabular}{l c c}        
\hline\hline                 
Parameter & Value & Units \\    
\hline                        
Primary mirror (M1) diameter   & 50.0 & m \\
Primary focal length      & 17.5 & m \\
Secondary mirror (M2) diameter & 12.0 & m \\
Back focal distance       & 14.0 & m \\
Tertiary mirror (M3) diameter  & $6 \times 8.6$ & m\\
Focal surface (FS) diameter    & 4.7 & m\\
\hline                                   
\end{tabular}
\label{tab:optical_parameters}      
\end{table}

The AtLAST optical configuration -- in particular, the Ritchey-Chr\'etien design with a short focal length -- is what enables the $2^\circ$ geometric FoV, while the rotating M3 allows AtLAST to select among at least six large instruments (Sect. \ref{sec:rxcabin}), exceeding the initial minimum requirements (Table~\ref{tab:design}). This design choice also reduces the number of mirrors to fewer than four, which adds the advantage of reducing the optical loading in the instrumentation.
\cite{Gallardo2024} presented the full geometric optical design for AtLAST, while \cite{Puddu2024} analyzed the beam intensity and cross-polarization patterns using physical optics calculations that include the effects of diffraction from the gaps in the primary mirror and scattering off of the secondary mirror support structure.

\begin{figure}[tbh]
    \centering
    \includegraphics[width=\hsize]{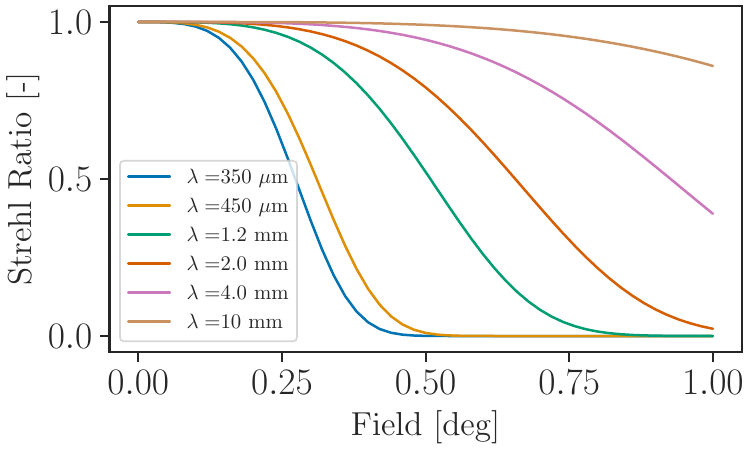}
    \caption{Strehl ratios vs.\ field angle for wavelengths of 350, 450, microns, 1.2, 2.0, 4.0 and 10 millimeters. We note this are the uncorrected Strehl ratios, native to the telescope optics, and do not consider additional corrective optics in the receivers. Corrections that recover a larger FoV, particularly at shorter wavelengths, are considered in \cite{Gallardo2024}.}
    \label{fig:strehlratios}
\end{figure}

We must note here that the $2^\circ$ geometric FoV of the design does not ensure a $2^\circ$ diffraction-limited FoV for the full range of frequencies that AtLAST will cover.
Two-mirror telescopes like AtLAST are in general able to cancel two of the three first-order optical aberrations, leaving astigmatism uncorrected.\footnote{In this context, we consider AtLAST optically to be a two-mirror design. The folding tertiary mirror (M3) of AtLAST is flat, and therefore does not enter the considerations for the correction of aberrations.  Options for a shaped M3 were considered early in the AtLAST design study, but were found to introduce an unwanted elevation dependence in the beam shape.} The presence of astigmatism limits the diffraction-limited FoV, in the absence of corrective optics. The result can be seen in Fig.~\ref{fig:strehlratios}, which shows the uncorrected Strehl ratios for several representative frequencies.  A Strehl ratio $\geq 0.8$ is considered diffraction-limited, so it is clear that, without corrections, the useful FoV AtLAST would be quite limited at high frequencies.

Fortunately, it is possible to correct astigmatism within the science instruments themselves by introducing an asymmetric biconic optical element (e.g., a lens) in the optical chain. The optimal place to insert this asymmetric optical element is the image of the entrance aperture (Lyot stop) or a neighboring optical surface. This approach has been developed and presented elsewhere \citep{Lou2020, Gallardo2022, Huber2022}, and has been implemented in various (sub)millimeter instruments \citep[e.g.,][]{Dicker2018, Gallardo2022, Huber2024, Gallardo2024cmbs4}.  \cite{mroczkowski2023} showed preliminary results with corrective optics, while \citealt{Gallardo2024} presents the full camera design concepts for optics within the instrument that correct much of the astigmatism and help recover significant portions of the geometric FoV of the telescope.

\section{AtLAST antenna structure}\label{sec:antstruct}

In this section, we discuss the overall antenna structure and how it achieves the design requirements.
\cite{Reichert2024} presents the technical flow-down {from the} requirements to the final design concept for AtLAST from an engineering perspective, along with a more complete discussion of the structure and FEA. Here, we summarize the salient points with the astronomical observer user community in mind.

\subsection{Structural design approach}\label{sec:design_approach}

\begin{figure*}[h]
        \centering
        \includegraphics[width=0.85\hsize, trim={3cm 1.7cm 3cm 6cm},clip]{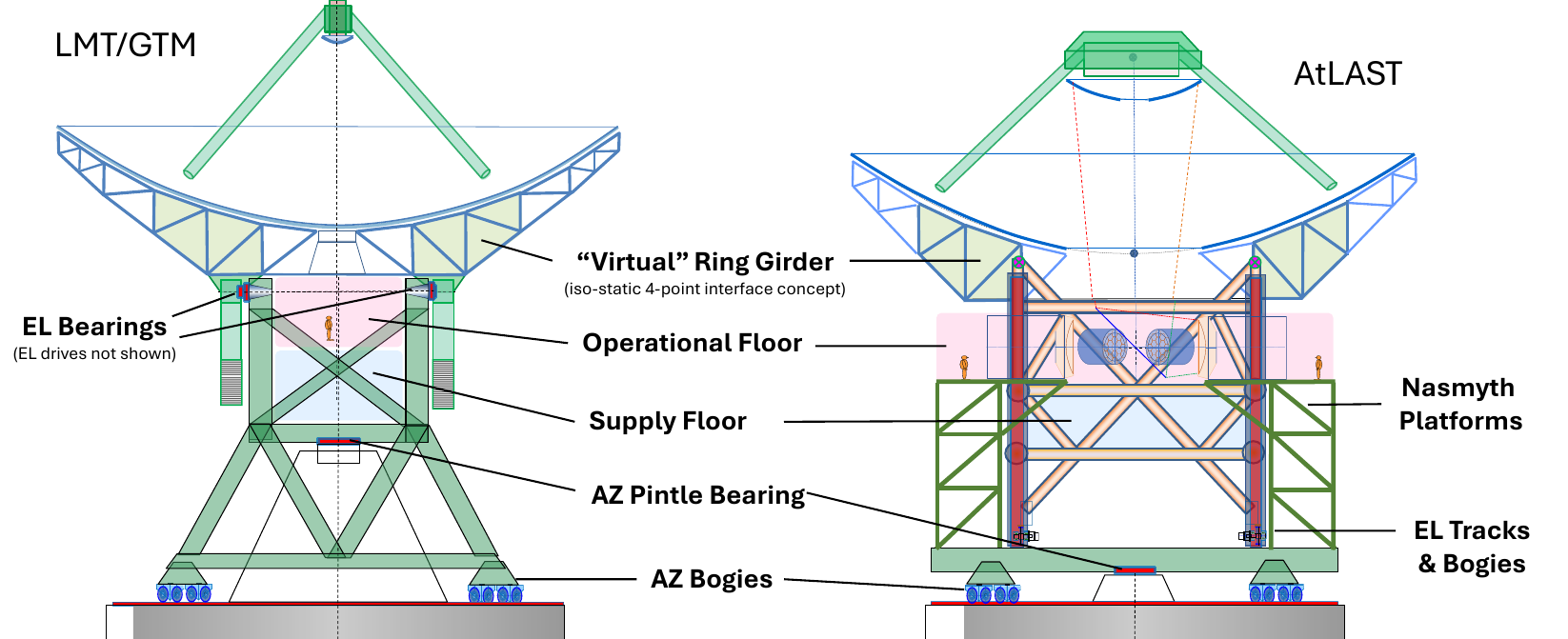}
        \caption{
        Sketches comparing the design concepts for the LMT/GTM 50-meter and AtLAST 50-meter. 
        {\it Left:} 
        Sketch of the elevation bearing concept for the LMT/GTM. 
        {\it Right:} Sketch of the bearing concept for AtLAST, allowing a greatly expanded volume for the receivers over more conventional designs.}\label{fig:bearing_sketch}
\end{figure*}

\begin{figure}[h]
        \centering
        \includegraphics[width=0.35\hsize, angle=0]{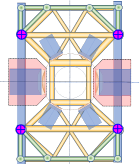}
        \includegraphics[width =0.62\hsize]{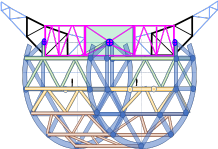}
        \includegraphics[width =0.62\hsize]{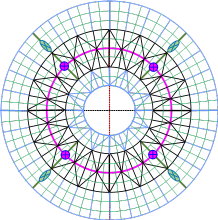}
        \caption{Sketches of the concepts for the instrument layout, elevation wheel, and backup structure. 
        {\it  Upper left panel:} Arrangements of the six science instruments inside the ``rocking chair'' elevation wheel.  The instruments are represented by two larger and four smaller blue rectangles. The two larger science instruments are mounted on the Nasmyth platforms (pink). The isostatic four-point interface between the main reflector backup structure and the rocking chair can be seen as the magenta circles with blue $+$ signs.  
        {\it Upper right panel:} Isometric view of the rocking chair elevation wheel. The isostatic four-point interface (magenta circles with blue $+$ signs) between the main reflector backup structure and the rocking chair elevation wheel can be seen more clearly.
        {\it Lower panel:} Bottom view of the backup structure showing the isostatic four-point interface to the backup structure.
        }
        \label{fig:structure_gen}
\end{figure}

The final choice for the optical layout, as is described in this work in Sect.~\ref{sec:final_optical_design} and in~\cite{AtLAST_memo_1}, had a significant impact on the structural and mechanical design of the telescope mount. 
Previous designs for large (sub)millimeter telescopes, such as the 50-m LMT/GTM (see left panel of in Fig.~\ref{fig:bearing_sketch}), were optimized optically  for observations using a single beam or small focal plane array, and thus have limited FoVs. 
The request for a 2 deg FoV for AtLAST resulted in an optical design \citep{AtLAST_memo_1, Gallardo2024} with a very large (12~m) secondary reflector, as well as a very large (6~m by 8.6~m) tertiary housed in the receiver cabin. 
The required space for the receiver cabin would interfere with the location of the elevation bearings in {a} conventional design like that of the LMT.
      
An innovative solution was found by switching to a rocking chair design for the elevation structure. We note that our approach is similar to the approaches taken by the designs of upcoming extremely large optical telescopes like the European Extremely Large Telescope \citep[ELT; ][]{Spyromilio2007, Tamai2014ELT} and the Giant Magellan Telescope \citep[GMT;][]{Johns2012}.
The rocking chair elevation wheel configuration {allows one} to expand the receiver cabin space available for the tertiary mirror and Cassegrain-mounted instruments.
Large openings in the center of the elevation wheel provide ample space for two Nasmyth platforms, placed outside the elevation rotating structure; these platforms  in turn support two large science instruments (receivers) that remain stationary in elevation (see right panel of Fig.~\ref{fig:bearing_sketch} and the left upper panel of Fig.~\ref{fig:structure_gen}).  
The four smaller Cassegrain-mounted receivers are attached to the top of the elevation wheel structure and rotate together in elevation as the telescope moves and points.  The receiver size allocations and cabin configuration and dimensions are discussed further in Sects.~\ref{sec:rxcabin} and \ref{sec:inst_sizes}, respectively.

The design of the telescope back-up structure (BUS) follows the so-called isostatic four-point design principle, introduced in the 1970s for the design of large radio reflectors \citep{BaarsKaercher2018}, for the interface with the main reflector BUS. 
The four-point principle ensures the isostatic decoupling between those two structural subsystems, introducing homology to the design, which in turn avoids first-order astigmatism and coma and reduces the relative gravitational deformations of the reflector by an order of magnitude.
The proportions of the elevation wheel are adapted to the four points of the BUS (see upper right and lower central panel of Fig.~\ref{fig:structure_gen}), which ensures isostatic decoupling between the BUS and the elevation wheel \citep{KaercherBaars2014, BaarsKaercher2018}. 

Throughout the design process, we chose to rely as much as possible on industry-standard, cost-effective materials such as steel and aluminum for the large scale structures, and to use materials with optimized mechanical properties such as carbon fiber reinforced plastics (CFRP) only when compensation by a metrology system is not possible or practical. Exceptions to this include the primary reflector panel segments and the secondary reflector, which both require the use of CFRP to reduce thermal deformations and weight \citep[see][]{Reichert2024}.

\begin{figure*}[bth]
    \centering
    \includegraphics[width=0.495\hsize]{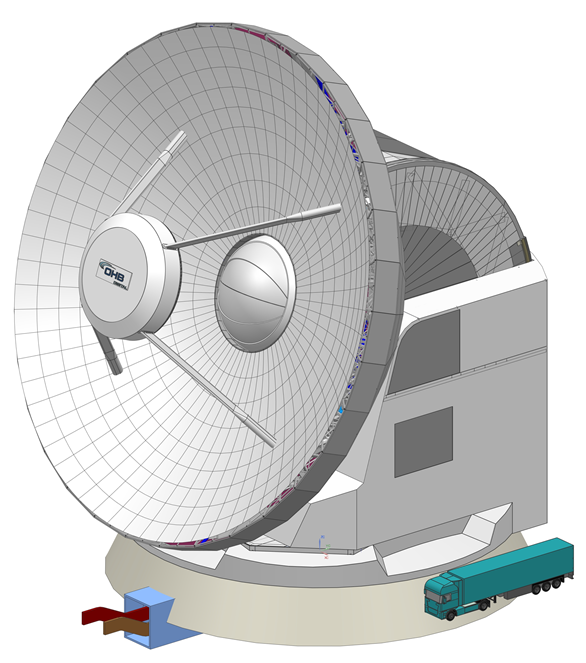}
    \includegraphics[width=0.495\hsize, trim={0.0cm 0.8cm 0.8cm 0.0cm},clip]{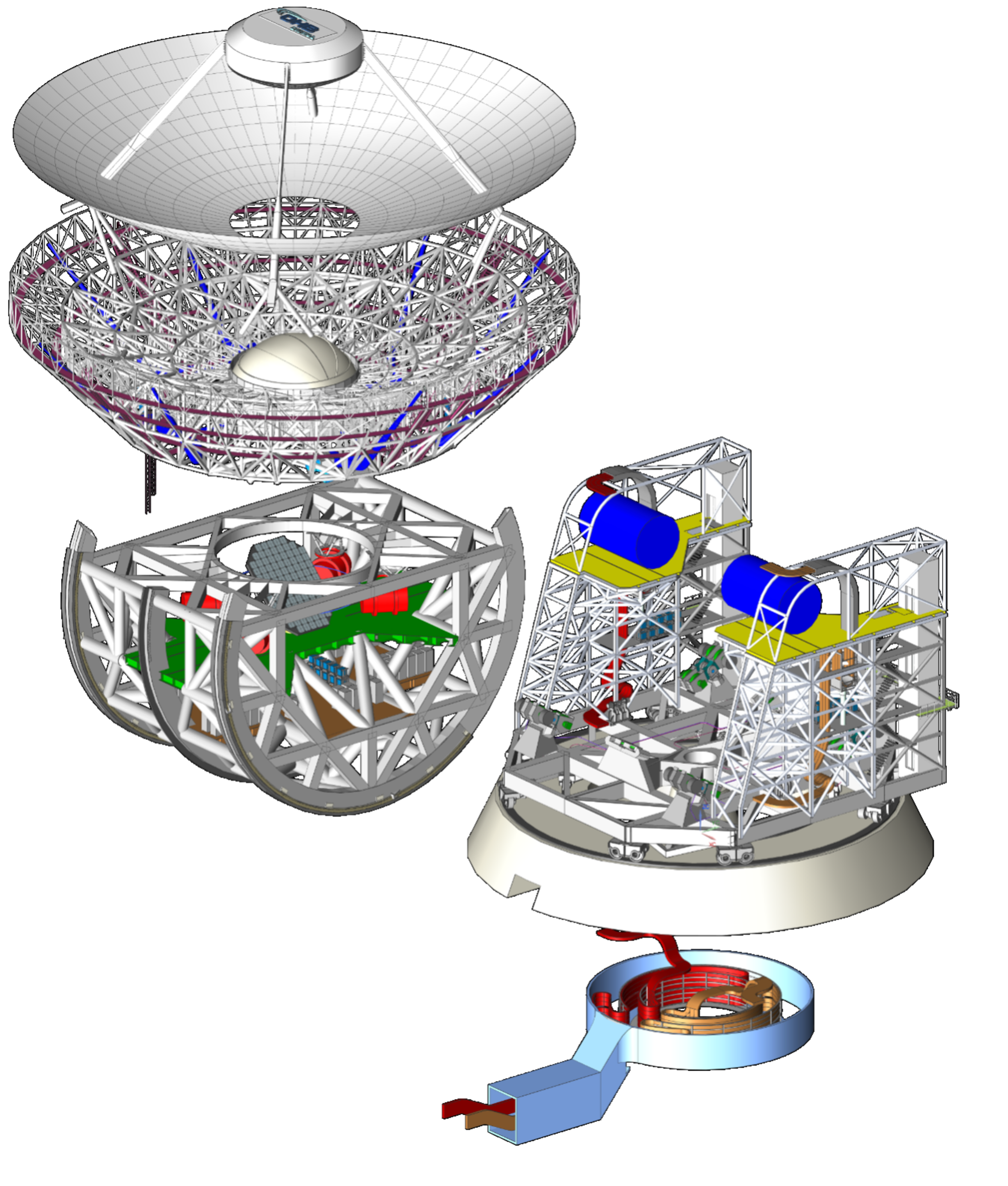}
    \caption{Internal and external views of the AtLAST concept CAD model. {\it Left:} Rendering of the complete telescope, using the AtLAST CAD model.  A truck is shown for scale. 
    {\it Right:} Exploded view of AtLAST that highlights the internal components. The exploded view shows, from top to bottom, M2, M1, the backup structure, the elevation rotating structure, (offset to the right) the azimuth rotating structure, and the underground cabling structure that is part of the azimuth cable wrap.
    The Nasmyth platform is highlighted in yellow here, and the Cassegrain platform is highlighted in green.}
    \label{fig:atlast_cad_model}
\end{figure*}

We verified the intended deformation behavior through FEA of the concept described in Sect.~\ref{sec:achieving_perf}, and provide detailed outputs of the FEA in the online data \href{https://zenodo.org/records/14673697}{available on Zenodo}.
The final accuracy of the passive reflector under gravity loading is better than 200~$\mu$m RMS over the whole range of motion, from 20$^\circ$ to 90$^\circ$ in elevation.
{Larger deviations ($>200~\mu$m) are expected to occur only in small areas around the isostatic four-point supports.
The gravity deformations due to changes in elevation are entirely repeatable, and our finite element modeling indicates that the gross gravity deformations can be largely corrected \citep{Reichert2024}. Further refinement can be obtained through empirical determination of the gravitational deformations, which would then inform the structural gravity compensation model (as is standard practice for other large single-dish facilities). }

Slow transient deformations due to temperature changes or wind deformations on timescales of ten seconds to thousands of seconds can be observed by sensors and corrected in a closed-loop control system that drives actuators for compensation. 
Deformations due to temperature effects are further minimized by external cladding and forced ventilation inside the backup structure and the elevation wheel, similar to the Institut de Radioastronomie Millimétrique 30-meter Telescope (IRAM 30-m; \citealt{Baars1987}) and the 50-meter LMT/GTM \citep{KaercherBaars2000}. Temperature sensors inside the backup structure and the rocking chair can be used for closed-loop corrections to improve the response to thermal deformations.
As an alternative to the active compensation approach, one might consider the large-scale use of materials with very low coefficients of thermal expansion and a very high ratio of stiffness to weight (e.g., CFRP). {However, from a financial perspective, we exclude this option due to the higher associated cost. Furthermore, there is a risk that CFRP will not achieve the required passive performance, meaning that closed-loop corrections will be required regardless of material choice.}
Steady-state deformations due to wind loading under 10~m~s$^{-1}$ wind speed are less than 25~$\mu$m RMS, mainly due to strong astigmatism. Active corrections are only required for very low-order Zernike polynomials. The dynamic wind effects on the reflector shape are an order of magnitude smaller than the steady-state effects and are only relevant for pointing. 

During fast scanning, depending on the scanning pattern, the intervals of high acceleration need to be minimized, since the surface accuracy cannot be maintained by active corrections (closed control loop between sensors and actuators) for short, transient disturbances although they might be predictable. The frequency content of the scanning trajectories will be defined by the axes control unit to be below the frequency range of the first natural modes of the structure (control of jerk). {For example, in the relatively demanding Lissajous daisy scan strategy case shown in Fig.~\ref{fig:liss_scan}, the frequencies of the accelerations are $\lesssim 0.1$~Hz, which is much lower than the natural frequencies ($\gtrsim 1.8$~Hz), discussed later in Sect.~\ref{sec:eigenfreq}, and can be expected not to excite resonances.}

Two coupled metrology systems for active corrections by the closed-loop control systems are foreseen to compensate slow, repeatable and non-repeatable, but observable deformations (e.g.,\ transient thermal deformations and the low-frequency component of wind loads). One system will be dedicated to the primary reflector surface and the maintaining of the optical alignment of the primary, secondary, tertiary mirror and the instruments (see Sect.~\ref{sec:ActiveOptics_and_Alignment}). A second system (termed as flexible body compensation (FBC)) will address structural deformations outside the optical path that affect the pointing accuracy of the telescope  (see Sect.~\ref{sec:pointing_accuracy}).

\subsection{AtLAST computer-aided design model}\label{sec:CAD}

The full computer-aided design (CAD) model for AtLAST is shown in Fig.~\ref{fig:atlast_cad_model}, with an exploded view in the right panel in order to show how several of the major assemblies come together. 
In Fig.~\ref{fig:atlast_cad_model}, one can see the 50-m primary mirror and 12-m secondary mirror, which are carried along with the rocking chair elevation rotating structure (Sect.~\ref{sec:ERS}).  The rocking chair is cradled on the azimuthal support structure, carried by two nested azimuthal tracks. Also carried by the azimuthal support are the Nasmyth platforms and the housing for the receiver cabin discussed in Sect.~\ref{sec:rxcabin}. The shaded light blue assembly, representing the cable wrap and power housing, will be located underground. Mass estimates for the major components as well as the receivers, discussed in Sect.~\ref{sec:insts}, are provided in Table~\ref{tab:masses}.

\begin{table}[tbh]
\caption{Mass estimates for various telescope components and equipment.}
\centering                          
\begin{tabular}{l c}        
\hline\hline                 
Structure or Component & Mass (tons) \\    
\hline                        
Telescope structure total                    &  4400 \\ 
Azimuthal structure (w/o instr.)             &  1390 \\
Elevation rotating structure (incl. ballast) &  3000 \\
Main reflector, backup structure, subreflector            &  1080  \\
AZ Wheel load                        &  $\leq$200 \\
Nasmyth-mounted instruments             &  2$\times$30 \\ 
Cassegrain-mounted instruments         &  4$\times$10 \\
\hline                                   
\end{tabular}
\tablefoot{The instrument mass estimates listed are the maxima allowed for those mount points.}  \label{tab:masses}      
\end{table}

\subsection{AtLAST finite element modeling}\label{sec:fem} 
The key drivers for the structural design of the AtLAST telescope are the surface accuracy and pointing accuracy requirements. As is customary in antenna design, these requirements were distilled into error budgets for each engineering consideration. A finite element model was built and used to verify the budget allocations of the passive structure due to environmental loads (gravity, wind, and temperature). Here, we use the term “passive structure” to refer to that before any active compensation of unwanted deformations by a closed-loop control system. 

\begin{figure}[tbh]
        \centering
        \includegraphics[width=0.99\hsize]{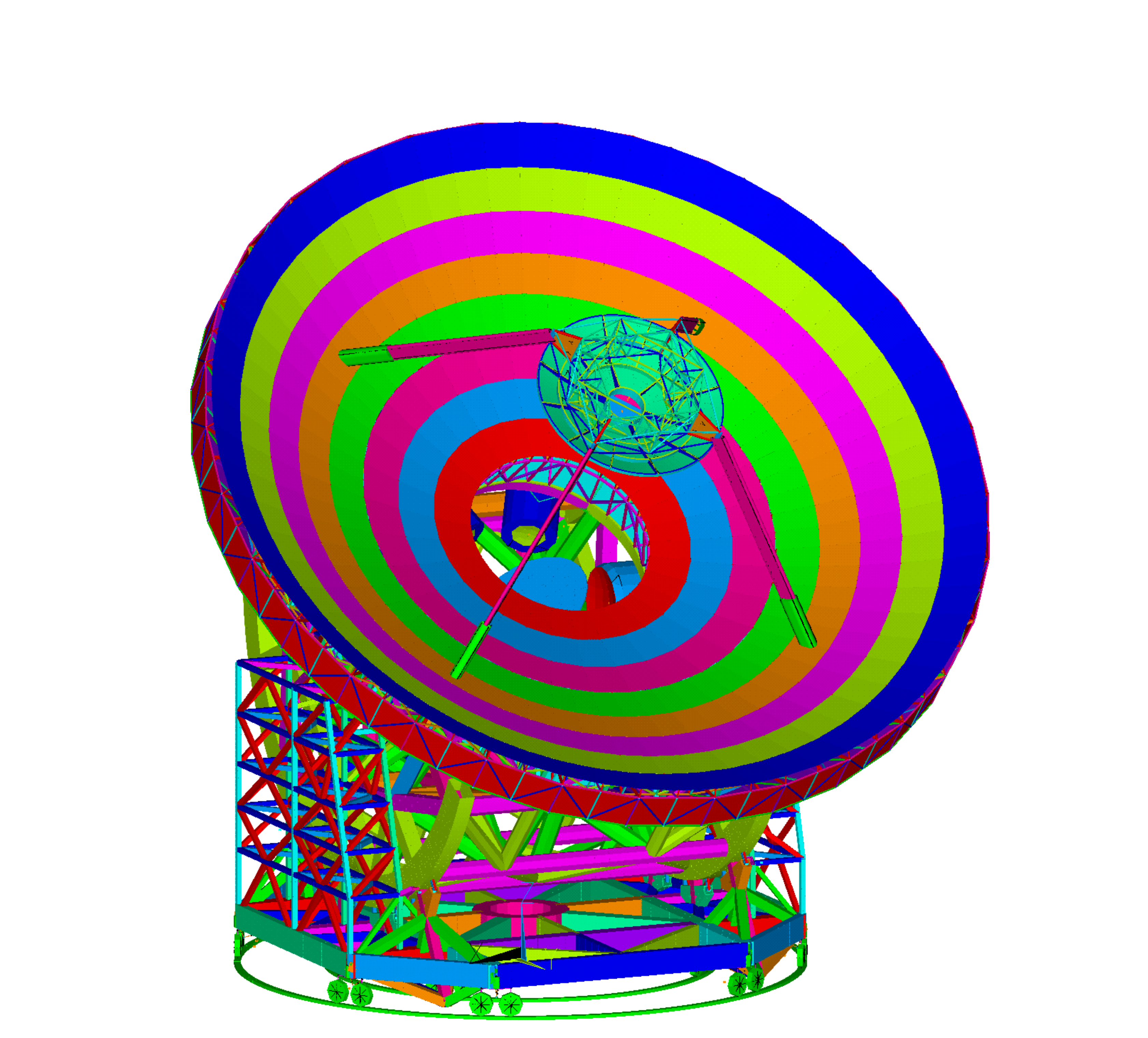}
        \caption{Overall telescope finite element model for AtLAST. Each color represents a specific, mechanical cross section or shell section property.}
        \label{fig:fe_model_total}
\end{figure}

\begin{figure}[tbh]
        \centering
        \includegraphics[width =\hsize]{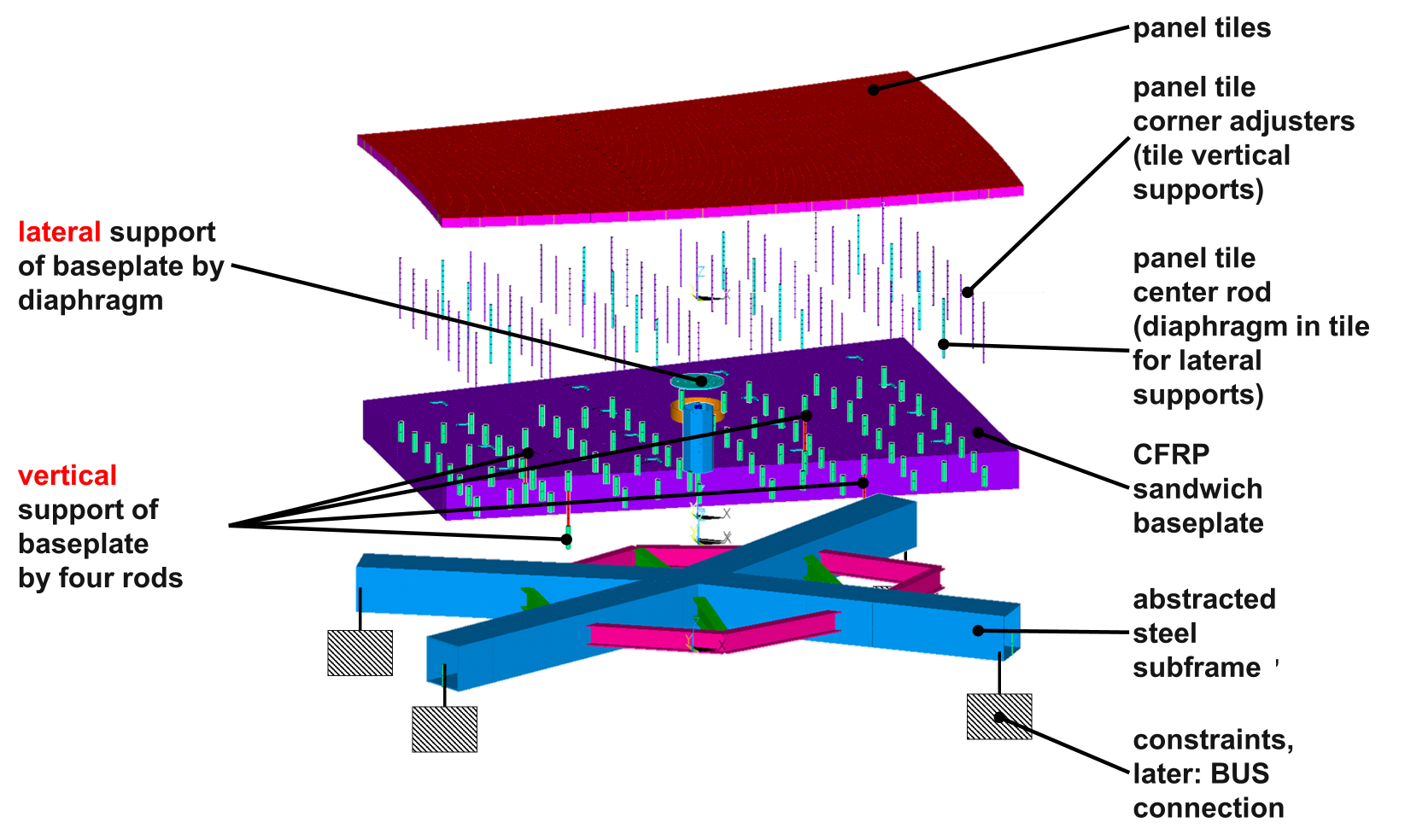}
        \caption{Explosion of a simplified FE panel segment model.  This segment is approximately 2.8 meters in length, and represents just one of the $\approx 400$ segments in the M1 design. The FE model here is used for evaluation of contributions to the surface error budget.}
        \label{fig:fe_model_panelseg}
\end{figure}

For the modeling and the evaluation, we used the Ansys$^\textsc{TM}$ 2021 R1 software suite.\footnote{\href{https://www.ansys.com/}{https://www.ansys.com/}}
The FEA was done using two analysis submodels. The first is the overall telescope model (see Fig.~\ref{fig:fe_model_total}) comprising the major load carrying structural elements with simplified panel segments to analyze error contributors allocated to the primary backup structure and overall structure. 
The second is a detailed model of the panel segments (see Fig.~\ref{fig:fe_model_panelseg}) that aims at the design of a panel segment to fulfill the surface error budget allocation.

\subsubsection{Primary mirror surface}\label{sec:primary}
The main reflector consists of eight rings of panel segments (see the online data \href{https://zenodo.org/records/14673697}{available on Zenodo}). Each panel segment consists of a pattern of 10-16 tiles of machined aluminum, depending on location in the primary. The arrangement of tiles for a specific panel rings depends on the maximum tolerable edge length of the panel tiles. 
This value is driven by the material chosen and the required machining precision, which depends on the absolute part size. For instance, because the part heats up during milling, thermal deformations become manufacturing errors. In addition, flexible deformations due to milling forces also depend on clamping conditions and can lead to manufacturing errors.  For these reasons, we chose a tile size of 0.7~m for the design.
Using larger panel tile sizes requires research on alternative materials, presumably composite materials, and their manufacturing and cost. In a preliminary trade-off, the choice was made to use aluminum machined panel tiles like those in existing radio telescopes (e.g., on IRAM's  Northern Extended Millimeter Array, or NOEMA) since composite materials would incur higher costs.

Due to the desire to allow solar observations (see Sect.~\ref{sec:solar}), we estimate a temperature differential $\Delta T \simeq 70-100^\circ$~C in the range of potential temperatures experienced by the panels in the primary reflector.  For aluminum, we expect a linear expansion of 1.7~mm per tile, and we designed the gaps to be 3~mm in order to provide a safety margin for thermal expansion and contraction.  We note that this gap size is slightly larger than the typical 1-2~mm gaps for smaller, 6-meter aperture (sub)millimeter telescopes \citep{FluxaRojas2016, Gudmundsson2021}. As in \cite{FluxaRojas2016, Gudmundsson2021}, we simulated the impact of the gap size on the beam shape using the physical optics package Ticra GRASP\footnote{\href{https://www.ticra.com/software/grasp/}{https://www.ticra.com/software/grasp/}}. In \cite{Puddu2024}, we show that gap sizes in the 1--5~mm range are subdominant to Ruze scattering and do not significantly affect the beam sidelobes, justifying our design choice.

The panel tile array is supported by a dedicated panel segment frame. A typical model of a panel segment frame can be seen in Fig.~\ref{fig:fe_model_panelseg}. The panel tiles are mounted and adjusted on a CFRP sandwich baseplate, which is mounted to a steel subframe by four normal (to the mean panel surface) adjuster rods and a thin metal diaphragm that takes the panel's in-plane (lateral) loads (e.g., when the reflector is oriented in low elevation angles). The diaphragm is located in the center of gravity of the assembly, which consists of panel tiles, panel tile adjusters, and a CFRP sandwich baseplate to avoid the introduction of bending moments by lateral gravity reactions. In this regard, the design principle is similar to a classical optical mirror support. 
The CFRP baseplate serves the function of decoupling thermal deformations of the steel backup structure and subframe from the panels. 

The panel segment's steel subframe is mounted to the main reflector steel backup structure at the panel segment's corners by a set of four longitudinal actuators in a statically overdetermined way, normal to the mean panel segment surface (one normal translational degree of freedom (DOF), two rotational degrees of freedom  (DOFs)). Additionally, there are three actuators controlling the panel segment's in-plane translation (two DOFs) and rotation (one DOF). Four panel segments share one actuator for the normal adjustment direction. 
The intention is to constrain the four panel corners in normal direction to follow the global backup structure deformations 
(which are then compensated by the actuators in a closed control loop) without generating offsets between the edges of the panel segments. The alternative to this approach is to resort to a static fully determined support of a panel segment in six DOFs where edge sensors are required between each set of neighboring panel segments.
The latter alternative has the downside of an overwhelmingly large sensor number (for a structure that is constantly fully exposed to the natural environment) and resulting control effort, and is therefore not considered as a baseline solution.

\begin{figure}[tbh]
        \centering
        \includegraphics[width =0.9\hsize, trim={14.0cm 0.5cm 16.0cm 10.0cm},clip]{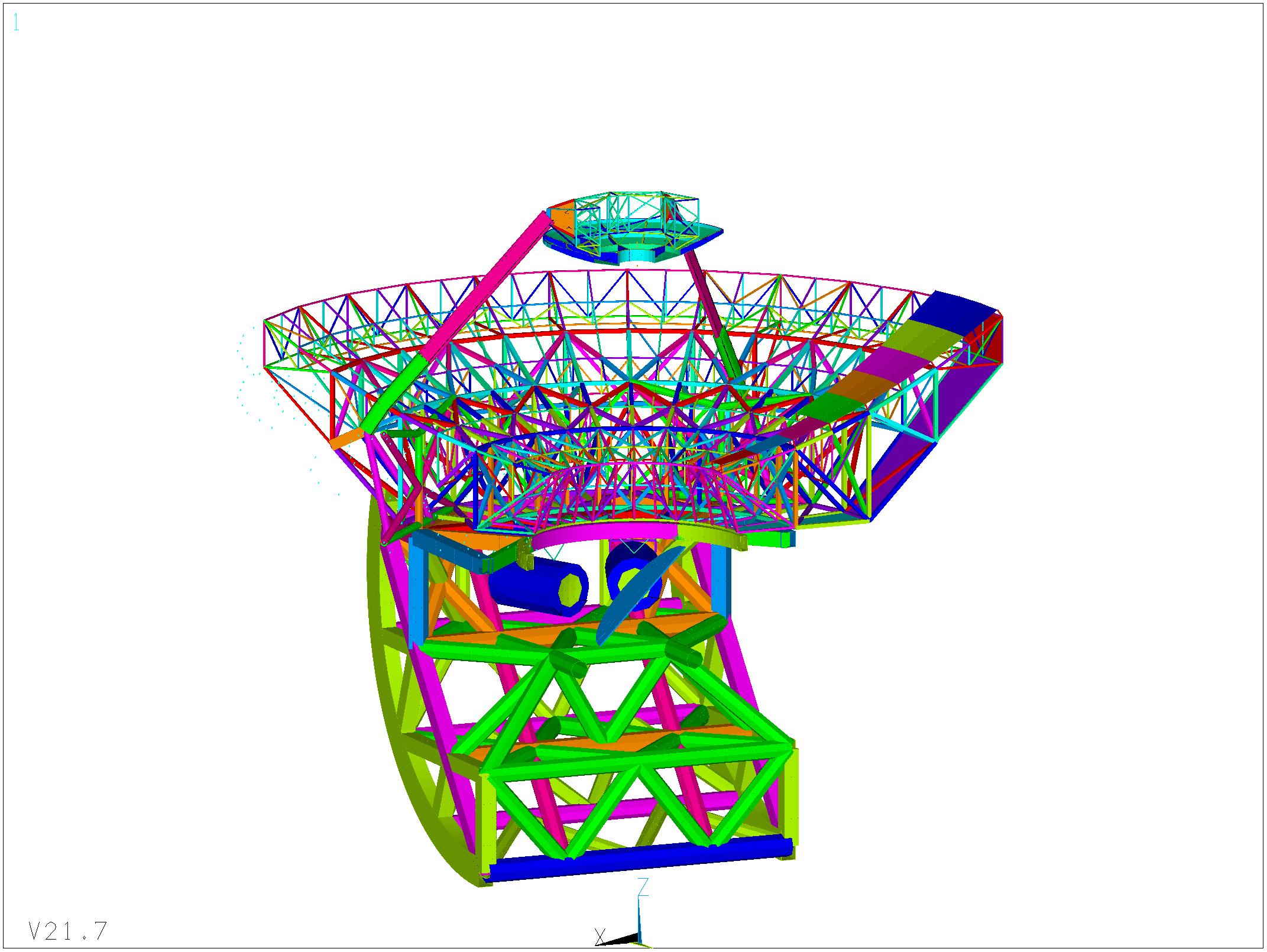}
        \caption{Cut-through view of the finite element model for elevation rotating structure.  The two dark blue cylinders are 2.6 meters in diameter and represent two of the four potential Cassegrain-mounted instruments, which corotate in elevation.  The cyan half-disk in the center represents (half of) the surface of M3. Each color represents a specific, mechanical cross section or shell section property.}
        \label{fig:fe_model_erswedge}
\end{figure}

\subsubsection{Elevation rotating structure}\label{sec:ERS}

The elevation rotating structure comprises a rocking chair structure that provides a large support base for the main reflector and leaves space around the elevation axis to host massive (30 ton; see Table~\ref{tab:masses}) Nasmyth instruments. As an elevation bearing, the rocking chair provides two skids whose tracks are supported by bogies on the azimuth rotating structure which react to radial and lateral forces.
The best arrangement of bogies for a robust centering of the elevation axis and the feasibility of a very precise yet robust track that resists high Hertzian stress is still under investigation, the details of which are reported in \cite{Reichert2024}. The approach intends to avoid hydraulic bearings due to their maintenance effort and operational costs at the geographic altitude at which the telescope will be constructed.  
As is shown in Fig.~\ref{fig:fe_model_erswedge}, the elevation rotating structure consists of all components rotating about the elevation axis. The rocking chair structure to which the elevation tracks are fixed enables rotation about the elevation axis by sitting on bogies.

A cut through the elevation structure is shown in Fig.~\ref{fig:fe_model_erswedge}.
The elevation rotating structure is currently expected to weigh 2810 tons with some trim mass for later auxiliary components already included. The goal is not to exceed 3000 ton in the course of the ongoing development (see Table~\ref{tab:masses}).

\begin{figure}[tbh]
        \centering
        \includegraphics[width =0.95\hsize, trim={12.0cm 6.0cm 12.0cm 10.0cm},clip]{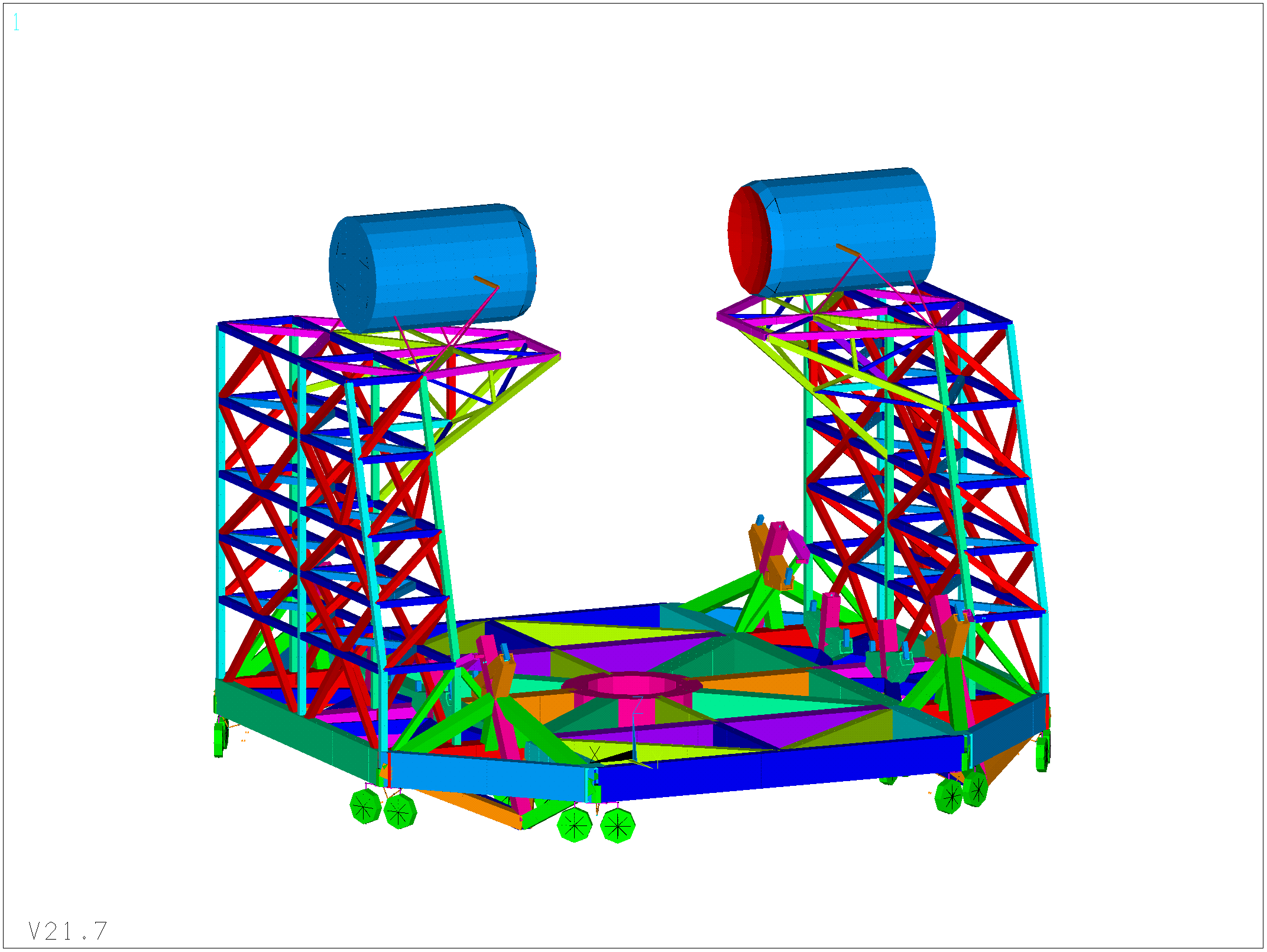}
        \caption{Finite element model of azimuth rotating structure.  In this, the two large blue cylinders (4.6 meters in diameter) represent possible monolithic Nasmyth instruments that are fixed in elevation.}
        \label{fig:fe_model_ars}
\end{figure}

\subsubsection{Azimuth rotating structure}\label{sec:ARS}

The azimuth rotating structure is a wheel-on-track design based on the design concepts used in the 64-m SRT  and in the 50-m LMT/GTM. It comprises:
\begin{itemize}
    \item twelve bogies, with two wheels per bogie, riding on on two track rings.
    \item the azimuth support truss supported by these bogies.
    \item the elevation bogies (lateral+radial support) carried by the azimuth support truss.
    \item the Nasmyth instruments racks or towers, also carried by the azimuth support truss.
\end{itemize}
The azimuth rotating structure can be seen in Fig.~\ref{fig:fe_model_ars}. The bogies (small green wheels shown in Fig.~\ref{fig:fe_model_ars}) run on seamless, welded tracks, which is similar to the approach chosen for the LMT/GTM. One can also see in Fig.~\ref{fig:fe_model_ars} the approximate package volumes of the Nasmyth instruments as blue cylinders on top of the racks or towers, and are supported by (motorized, controlled) hexapods, which provide stable support with six DOFs for translational and rotational adjustments.

\subsubsection{Eigenfrequencies and mode shapes}\label{sec:eigenfreq}

The mechanical natural frequencies, or eigenfrequencies, of the structure are an important property as they represent a limitation for the control loop bandwidth of the main axes drives. The controllers' bandwidths limit their ability to reject environmental disturbances on shorter timescales (higher frequencies) such as wind gusts. 
A mechanical structure can be dynamically excited and will vibrate if the loading vector is similar to a mode shape and the excitation frequency is close to the natural frequency. For a sudden transient excitation with a broad frequency content (for example when considering a step function or a wind gust), the structural response will be a decaying oscillation consisting of a superposition of several excited natural mode shapes and frequencies. The natural mode shapes where a significant local deformation of the reflector panels can be observed (which could affect the surface accuracy) are usually at frequencies well above 10~Hz and in a range where wind spectra usually show low excitation magnitudes. The impact of lower frequency modes on the surface accuracy, where deformations in the BUS can be observed, are covered by the steady-state wind deformation analyses. Therefore, the major impact of the natural modes is the pointing or tracking stability due to wind gusts. Analysis of this error contribution is currently ongoing. A dynamic excitation by scanning trajectories can be minimized by controlling the jerk of the trajectory and the frequency content of the commanded acceleration. The investigation and analysis of scanning scenarios is also currently ongoing. Depending on the acceleration levels, the surface and pointing accuracy and thus the effective observation time when scanning might be compromised.
The natural frequencies are a function of the structure's ratios of stiffness to mass. Plots showing the mode shapes can be found in the online data \href{https://zenodo.org/records/14673697}{available on Zenodo}.
For the current design status, the first five eigenfrequencies can be seen in Table~\ref{tab:eigenfreq}. The first natural mode shape at 1.8~Hz is a twist of the M2 and crown on the quadripod. This mode can be probably improved but is not critical for the main axes control loops. The second and fourth mode shapes are more important for the main axes control loop bandwidth, and show reasonable frequencies given the size and weight of the structure to cope with excitation
by wind gusts and earthquakes.

\begin{table}[h]
\caption{Eigenfrequencies of the telescope structure}
\centering                          
\begin{tabular}{c c l}        
\hline\hline                 
Mode & $f_i$ [Hz] &  Shape description  \\    
\hline                        
\#1 & 1.8 & M2 twist about LOS on quadripod\\ 
\#2 & 1.9 & ERS longitudinal w.r.t.\ the ARS\\ 
\#3 & 2.1 & ERS lateral w.r.t.\ the ARS\\ 
\#4 & 3.0 & Full telescope on track rotation about AZ \\ 
\#5 & 3.1 & Lateral bending for Nasmyth instr.\\ 
\hline                                   
\end{tabular}
\tablefoot{Results of the modal analysis to determine the eigenfrequencies of the antenna structure.  The first five frequencies, $f_i$, where $i=1,2,3,4,5$, and descriptions of the mode shapes are listed for EL $= 90^\circ$.  Mode \#5 describes the lateral bending of the Nasmyth support platform due to the mass and inertia of the Nasmyth instrument.
Here, LOS stands for line of sight, ARS stands for azimuthal rotating structure, and ERS stands for elevation rotating structure.}  
\label{tab:eigenfreq}      
\end{table}

\subsection{Energy supply and power demand}\label{sec:energy}

In designing the energy system for AtLAST, determining the future demand of the new 50-m telescope can pose challenges. Both the integration of innovative instrumentation with yet-to-be-defined power requirements, and the need to drive such a large (4400~ton; see Table~\ref{tab:masses}) structure faster than similar structures, add complexity to the challenge. Presently, we estimate that the power demand, summarized in Table~\ref{tab:energy}, will include 500-700~kW for instrumentation electronics, 480~kW for cryogenic cooling, and an average demand of 500~kW for the telescope drives (namely, the motors), with peak loads for the drives reaching up to 1.7~MW. 

\begin{table}[h]
\centering                          
\caption{Expected peak energy demands for AtLAST}
\begin{tabular}{l c c}        
\hline\hline                 
Item & Value & Units \\    
\hline                        
Instrument Electronics         & 600 & kW\\
Cryogenic Demand of Nasmyth insts    & $160\times 2$   & kW\\
Cryogenic Demand of Cassegrain insts   & $40 \times 4$    & kW\\
Peak AZ Drive Demand           & 1100  & kW\\
Peak EL Drive Demand           & 600   & kW\\
Heating, ventilation, \& air conditioning                   & 200   & kW\\
\hline
Total Peak Demand              & $\approx 2980$ & kW \\
\hline                                   
\end{tabular}
\tablefoot{Rough order of magnitude expected peak energy demands for AtLAST. Of this, up to $85\%$ of the 1.7~MW peak drive will be recoverable, while AtLAST's energy recovery system is expected to also be able to shave the peak demand from the drives \citep[see][]{Kiselev2024}.}             
\label{tab:energy}      
\end{table} 

Recently, we have studied the feasibility of off-grid systems based on renewable energy sources to meet AtLAST's power needs on the Chajnantor plateau. {In addition to being considered more environmentally responsible in comparison to diesel-only setups, hybrid energy systems featuring a large photovoltaic array, lithium-ion batteries, and backup diesel generation can be anticipated to be significantly less expensive over the lifetime of the project \citep[see][]{Viole2023, Valenzuela-Venegas2023}}. Furthermore, hydrogen is gaining traction as an energy carrier in Chile, and is considered a suitable future substitute for backup diesel generation. We have therefore included hydrogen in the proposed energy system. 

As was noted earlier, the main drives (AZ and EL) are expected to be the main sources of power consumption. Due to the high inertia of the full telescope structure, combined with a high acceleration requirement, considerable power peaks will occur during the acceleration and deceleration phases. During acceleration, the power peak is present as a high power demand with a flow direction from the grid to the drives. On the other hand, during deceleration the drive system generates power, which is typically converted to heat in corresponding brake resistors in most large telescopes. 

\begin{figure}[tbh]
        \centering
        \includegraphics[width =\hsize]{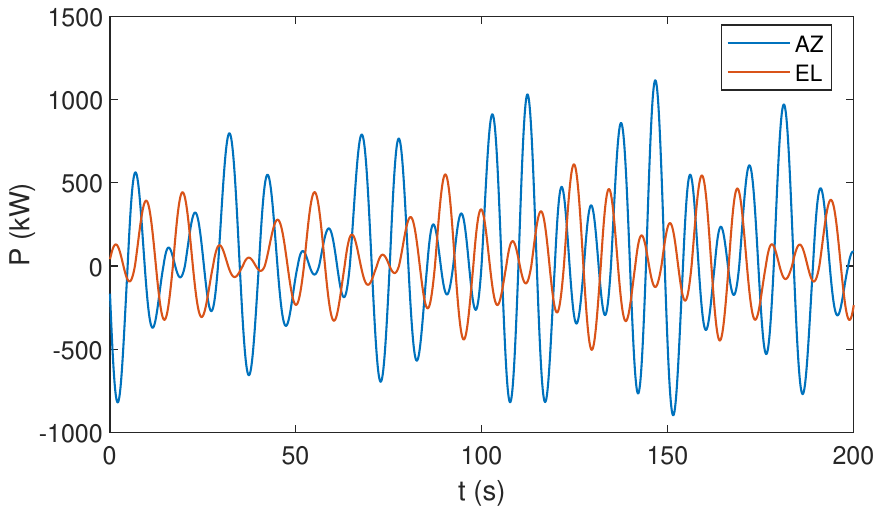}
        \caption{Power demand of AtLAST's main drives for the representative Lissajous daisy scan shown in Fig.~\ref{fig:liss_scan}. Taken together, the power demands of the scans represent a significant fraction of the total power budget, motivating the need for the energy recovery system described in \cite{Kiselev2024}.}
        \label{fig:power_demand}
\end{figure}

Taking into account the considerable power demand of AtLAST's main drives, ``classic'' telescope power management is not acceptable. Assuming {the example Lissajous scan shown in Fig.~\ref{fig:liss_scan}, which represents an extreme but plausible case for AtLAST,} the expected peak electric power is $\approx 1100$~kW for the AZ-drive and $\approx 600$~kW for the EL-drive. A time-ordered plot of the power demand is shown in Fig.~\ref{fig:power_demand}.  The major part of this power is due to inertia during acceleration and deceleration, and is thus reusable. For this reason, an energy recovery system based on supercapacitors has been designed and modeled in \cite{Kiselev2024}. Unlike other such systems, this offers the ability to reduce (or ``shave'') the electric power peaks, reducing the power system demand and drastically improving the overall efficiency of the telescope's drive system, which is a major energy consumer; this energy would normally have to be dissipated as heat through braking, but instead can be largely recovered and reused.

\subsection{Environmental effects}\label{sec:environment}
The Atacama Desert is not immune to poor weather or strong seismic events such as earthquakes and volcanic activity. 
While lightning protection and wind limits ($>25~\rm m\,s^{-1}$), which necessitate stowing, have already been considered, the team is now working on the detailed modeling of the acceleration during earthquakes.  
As is noted in \cite{alma_memo_413}, \cite{alma_memo_418}, and numerous other ALMA reports, the strength of seismic events is greatly diminished in the Llano de Chajnantor region compared to that at observatories like ESO's Paranal Observatory and the future ELT, which are closer to the Chilean coast and Chilean fault line.
This work is ongoing and will continue into the next phase of the AtLAST project as the design approaches full subsystem engineering.  In addition, the team plans to perform detailed hydrodynamic simulations to more robustly model the impact of high winds speed and turbulence.

\section{Achieving the required surface and pointing accuracy}
\label{sec:achieving_perf}

\subsection{General approach} \label{sec:achieving_perf_general}
The science cases place very stringent requirements on the antenna design, considering the size of AtLAST and the exposure of its mechanical parts (especially M1 and M2) to the environment. Without compensation, structures of the size and weight of AtLAST will deform on scales that are orders of magnitude larger than the required accuracy of the optical surfaces and optical alignment. Therefore, the structural design of the reflectors and optical path components is an art of maintaining the correct shape of the optical geometry in the presence of deformations due to environmental loads (gravity, wind, thermal loads) to within the {RMS HWFE}  requirement. 

The approach to mitigate the deformations of the AtLAST telescope consists of two main steps:
\begin{enumerate}
    \item Design and FE analysis of the static structure, using structural engineering principles described in \cite{BaarsKaercher2018} to achieve a baseline level of performance. {This includes the use of open-loop lookup-table-based compensation for the static deformations.}
    \item Using sensors and actuators in a closed-loop control system to compensate for residual deformation beyond what can be achieved within the static structure.  This is expected to bring performance to a much higher level of accuracy. This approach is what is meant here when referring to the {``metrology system''} in the context of the AtLAST project.
\end{enumerate}
Due to its size, AtLAST is one of the first planned or operational (sub)millimeter telescopes where active, closed-loop compensation systems are mandatory to meet the demands on HWFE and pointing accuracy. For example, the primary reflector surface requires an active system to adapt to transient environmental loads on $1-10$ second timescales. 
\cite{Reichert2024} presents the technical flow-down from requirements to the final design concept for AtLAST from an engineering perspective, along with a more complete discussion of the structure and FEA.
A more detailed overview of the closed-loop compensation concepts considered for AtLAST can be found in Sect.~\ref{sec:metrology_concepts}. 
The FEA model was used to verify the performance of the static design as a basis for the closed-loop compensation systems described  in Sect.~\ref{sec:perf_fe_hwfe_m1}~\&~\ref{sec:pointing_accuracy}.

\subsection{Surface accuracy}\label{sec:optical_quality_metrics}

\begin{table*}[tbh]
\centering                          
\caption{Analyzed and expected sources of focussing and alignment errors}
\begin{tabular}{l c c c c c c}        
\hline\hline                 
{ Error type}  & Analyzed & Uncorr.\ &  Corrected  & Compensation & See section \\    
 &  & [$\mu$m]  & [$\mu$m] & type & number\\
\hline
\multicolumn{1}{l}{M1 surface RMS HWFE} \\
\hline                        
Gravity BUS         & Yes & 200 & 5    & LUT/CL & \ref{sec:perf_fe_hwfe_m1}, \ref{sec:ActiveOptics_and_Alignment}\\
Gravity panels      & Yes &   8 & 8    & None & \ref{sec:perf_fe_hwfe_m1}\\
Wind BUS (3.5~m/s)    & Yes &  1  & $<1$    & CL     & \ref{sec:perf_fe_hwfe_m1}, \ref{sec:ActiveOptics_and_Alignment}\\
Wind panels (3.5~m/s) & Yes &  1  & 1    & None     & \ref{sec:perf_fe_hwfe_m1}\\
Wind BUS (9~m/s)    & Yes & 10  & $< 10$  & CL     & \ref{sec:perf_fe_hwfe_m1}\\
Wind panels (9~m/s) & Yes &  2  & 2   & None     & \ref{sec:perf_fe_hwfe_m1}\\
Thermal BUS (night)& Yes  & 12  & 1  & CL   & \ref{sec:perf_fe_hwfe_m1}, \ref{sec:ActiveOptics_and_Alignment}\\
Thermal panels (night)& Yes & 2 & 2  & None & \ref{sec:perf_fe_hwfe_m1}\\
Thermal BUS (day)  & Yes & 50   & 5  & CL & \ref{sec:perf_fe_hwfe_m1}, \ref{sec:ActiveOptics_and_Alignment}\\
Thermal panels (day)  & Yes & 5   & 5 & None & \ref{sec:perf_fe_hwfe_m1}\\
\hline                        
\multicolumn{1}{l}{M2 surface RMS HWFE} \\
\hline                        
Gravity BUS          & Yes&   10 & 10 & None & \ref{sec:perf_fe_hwfe_m2m3}\\
Gravity panels       & No &    2 &  2 & None & \ref{sec:perf_fe_hwfe_m2m3}\\
Wind BUS (3.5~m/s)     & No &    1 &  1  & None     & \ref{sec:perf_fe_hwfe_m2m3}\\
Wind panels (3.5~m/s)   & No &    1  &  1  & None     & \ref{sec:perf_fe_hwfe_m2m3}\\
Wind BUS (9~m/s)     & No & 7    & 7  & None     & \ref{sec:perf_fe_hwfe_m2m3}\\
Wind panels (9~m/s)   & No & 2    & 2  & None     & \ref{sec:perf_fe_hwfe_m2m3}\\
Thermal BUS (night)  & No & 2    & 2 & None & \ref{sec:perf_fe_hwfe_m2m3}\\
Thermal BUS (day)    & No & 4    & 4 & None & \ref{sec:perf_fe_hwfe_m2m3}\\
Thermal panels (night)& No & 1   & 1 & None & \ref{sec:perf_fe_hwfe_m2m3}\\
Thermal panels (day)   & No & 5   & 5 & None & \ref{sec:perf_fe_hwfe_m2m3}\\
\hline                        
\multicolumn{1}{l}{M3 surface RMS HWFE} \\
\hline                        
Gravity BUS    & No &   3 & 3 & None & \ref{sec:perf_fe_hwfe_m2m3}\\
Gravity panels & No &   1 & 1 & None & \ref{sec:perf_fe_hwfe_m2m3}\\
Thermal (night)& No & 1   & 1 & None & \ref{sec:perf_fe_hwfe_m2m3}\\
Thermal (day)  & No & 1   & 1 & None & \ref{sec:perf_fe_hwfe_m2m3}\\
\hline                                   
\multicolumn{1}{l}{Tolerances and alignment} \\
\hline                        
Panel manufacturing tolerances      & No &   8 & 8  & None & \ref{sec:optical_quality_metrics}\\
Active optical alignment (3.5~m/s)  & No &   - & 5  & None & \ref{sec:perf_fe_hwfe_m2m3}, \ref{sec:ActiveOptics_and_Alignment}\\
Active optical alignment (9~m/s)    & No &   - & 12 & None & \ref{sec:perf_fe_hwfe_m2m3}, \ref{sec:ActiveOptics_and_Alignment}\\
\hline                        
\end{tabular}
\tablefoot{Overview of analyzed and expected sources of error in the focusing and alignment of the optical surfaces. 
Analysis here refers to FEA or end-to-end (dynamic state space) modeling.
LUT refers to look-up tables (i.e.,\ reproducible values that can be stored in the system and are often appropriate for gross corrections).
CL refers to closed-loop corrections expected to rely on metrology and measurements used as feedback in the control system.
We note that the error corrections and methods of compensation are based on expectations and requirements that will be further refined as the concept evolves to a construction-ready design.
``None'' implies that the passive structure provides sufficient performance.
We further note that M3 is in the receiver cabin enclosure, so no wind is expected and the thermal deformations are not expected to exhibit strong diurnal variations.
Bulk tilts, rotations, and displacements of M2, M3, and the receivers with respect to M1 are not shown here but were considered in the detailed internal error budgets and will be a topic of future study. The values reported in this table supersede those reported in \cite{Reichert2024}.} 
\label{tab:surferrors}      
\end{table*}

The HWFE provides a quantitative way of measuring the impact of the overall RMS surface and alignment errors, and is the main parameter used both in computing the beam efficiency (in the Ruze formula; see Sect.~\ref{sec:surface_acc_req}) and in the Strehl ratio \citep[See e.g.,][]{Parshley2018}.  The impact of the HWFE is discussed in further detail by \cite{Puddu2024}. 
 The HWFE describes the signal degradation at the focal surface of the instrument, taking into account disturbances that can affect an electromagnetic wave traveling through the entire optical path of the telescope. The wavefront error also allows a spatially resolved assessment of the signal degradation across the curved focal surface. We note that the local RMS surface error of a single reflecting surface is identical to its contribution to the HWFE.

The systems-engineering approach to achieve the requirement is to address error contributors in a dedicated error budget per performance category -- here, HWFE from aperture to sensor. One important component of the error budget is environmental loads. The environmental contributors to the HWFE and their magnitudes are listed in Tab.~\ref{tab:surferrors}. The magnitudes were either verified by FEA where possible (bottom-up approach) and/or estimated (top-down approach). The target for the future steps is of course to verify the estimated budget values either by analysis or prototype breadboard tests.

As is common practice, we adopted the HWFE as a metric when establishing our engineering error budget, which was used to set constraints on each subsystem or component by tracking how their tolerance and achievable precision impact the overall optical performance. While the full engineering error budget is beyond the scope of this work, we note that \cite{Reichert2024} provides a more detailed review of the error budget and we simply discuss aspects relevant to the AtLAST conceptual design here.

\subsubsection{Primary mirror HWFE finite element analysis}\label{sec:perf_fe_hwfe_m1}
One of the major contributors to the telescope's half wavefront error budget is the surface deformation introduced by the primary reflector backup structure and the panels. From an engineering point of view, the contribution of the backup structure and the panels can be assessed in separate FE models, since their contribution to the error budget can be expressed as root sum squares. 
We show the results of the FE modeling of the primary mirror deformations under different conditions in the online data \href{https://zenodo.org/records/14673697}{available on Zenodo}.
These online materials show the full panel segment models included in the global telescope model, where the half wavefront errors reported there are the root mean squared contributions from the backup structure (whose repeatable contributions can largely be compensated by the active surface), the global structure, and the panel segments.
 The mostly repeatable contribution of the gravity deformations lies at the extreme limits in elevation (namely, $EL=90^\circ$ and $EL=20^\circ$), where the fractional error contributed by the backup structure can still be compensated by the active surface. 
 The results also show the HWFE caused by a quasi-static wind load for a wind speed of $10~\rm m \, s^{-1}$ at $EL=20^\circ$ and, as a worst case, approaching from the front (which means the azimuth angle of attack is $AZ_{wind}=0^\circ$). 
 The results further show the HWFE for a temperature deformation mode that can be compensated by an active primary reflector surface. 
 And finally, they reveal a temperature gradient normal to the surface, within the panel tiles, that cannot be compensated by any control system considered so far. It should be noted, however, that the magnitude of the temperature gradient is a deliberately chosen assumption, and its real magnitude will have to be verified in future analysis or prototype measurements.

{
\subsubsection{Secondary and tertiary mirror HWFE}\label{sec:perf_fe_hwfe_m2m3}

The impact of gravity on the secondary mirror and support structure were analyzed in the FEA, while the finite element model does not yet include in detail the local structural deformations of the tertiary mirror and support.
The thermal deformations of M2 will be minimized by material choice (i.e.,\ CFRP) for the BUS.
The expected contributions to the RMS HWFE can be found in Table \ref{tab:surferrors}, along with the expected contributions from residual misalignments of the optical surfaces after active compensation.
}

\subsection{Pointing accuracy}\label{sec:pointing_accuracy}

\begin{table*}[tbh]
\centering                          
\begin{tabular}{l c c c c  c c c}        
\hline\hline                 
{ Pointing errors}  & Analyzed & Uncorr.\ &  Corrected  & Type of Compensation  &  Sect. \\    
  &  &  [\arcsec] & [\arcsec] \\
\hline                        
Gravity        & Yes & 100 & 1 & LUT/CL & \ref{sec:pointing_accuracy},  \ref{sec:flexible_body_compensation}\\
Wind (3.5~m/s)   & Yes & 0.5 & 0.2  & CL     & \ref{sec:pointing_accuracy}, \ref{sec:flexible_body_compensation}\\
Wind (9~m/s)   & Yes & 3   & 1.5  & CL     & \ref{sec:pointing_accuracy}, \ref{sec:flexible_body_compensation}\\
Thermal (night)& Yes & 5   & 0.5 & CL & \ref{sec:pointing_accuracy}, \ref{sec:flexible_body_compensation}\\
Thermal (day)  & Yes & 5   & 1 & CL & \ref{sec:pointing_accuracy}, \ref{sec:flexible_body_compensation}\\
\hline                                   
\end{tabular}
\caption{{Overview of analyzed and expected sources of error in mechanical pointing errors.
While it is anticipated that the LUT approach will be sufficient to correct gravitational deformations, we include ``CL'' for them since the closed-loop system will be available to correct residual surface errors not amenable to the LUT approach.
Abbreviations are as in Table \ref{tab:surferrors}.  
We note that the anticipated corrections are based on expectations and requirements and may evolve as the concept evolves to a construction-ready design.}
}             
\label{tab:pointerrors}      
\end{table*}

{
To achieve our mechanical pointing accuracy requirement of 2\farcs5, we plan to implement a static pointing error correction model (SPEM) combined with closed-loop control compensation techniques.
One important component of the error budget addresses the environment as the root cause for pointing errors. The environmental contributors to the pointing errors and their magnitudes are listed in Tab.~\ref{tab:pointerrors}. As for the surface errors (see Sect.~\ref{sec:optical_quality_metrics}), the magnitudes of the pointing errors were verified by FEA whenever possible and estimated based on past experience whenever FEA was considered beyond the scope of this initial design study. A more detailed study will be carried out in the next phase of the project.

Since no 50-meter-class (sub)millimeter telescope operating up to $\sim 950$~GHz and having the same pointing requirements as AtLAST exists yet, we are unable to quantify at this stage how well a SPEM will perform.  
This will be the subject of the next phase in the AtLAST design.
However, given the experience of the facilities discussed in Sect.~\ref{sec:pointing_req}, we expect the requirement of 2\farcs5 mechanical pointing accuracy will not be achieved solely through the use of a SPEM. 
Additional sensor data and advanced algorithms will be needed to compensate, as much as needed, for transient disturbances. 
This system is called FBC
 \citep{Kaercher1999, Kaercher2006, BaarsKaercher2018}, and will be discussed further in Sect.~\ref{sec:metrology_concepts} .  
The root causes for transient disturbances are generally the wind and thermal environment to which the unsheltered structure of AtLAST is exposed. We note that new approaches using machine learning algorithms that include data from both sensors and astronomical pointing may offer improvements. A description of one such approach is given in \cite{nyheim2024machine}. In that work, though the training data were limited, they note significant improvements in the computation time necessary to achieve similar accuracy as standard astronomically corrected pointing models.
To learn about the transient disturbances expected from the environment, two dedicated 24-meter weather towers were constructed (one each at the two prospective AtLAST sites) and outfit with high temporal resolution, two- and three-dimensional anemometers.  
The data collected there will be used to inform further the finite element modeling, which will be reported in future works as part of the next phase AtLAST design.
In addition to the FBC approach, we anticipate the impact of transient thermal disturbances will be largely preempted by the planned inclusion of insulation as well as thermal conditioning through the HVAC system.

Pointing corrections need to be considered in conjunction with the primary closed-loop control system for the active surface and the optical alignment, described in Sect.~\ref{sec:ActiveOptics_and_Alignment}. Since the alignment of the optical elements and the pointing direction are coupled, the pointing will be the secondary property to be controlled after the M1 surface control and alignment of the optical components.
The reason for the coupling between the optical element alignment and the pointing is that, when the optical elements were considered as rigid bodies, any movement in three translational and two rotational DOFs (rotation about the optical axis does not affect pointing) can alter the pointing direction. This can be calculated using numerical derivatives of the pointing direction with respect to small translations and rotations in each DOF individually (linearization of disturbances about the nominal arrangement) for each optical element. Thus, any correction in the optical alignment can be translated into pointing corrections that can be made by the main telescope axes.
The FBC (as a metrology system) for the pointing performance on the one hand, and the active surface and alignment system for the HWFE on the other, can be physically distinguished by the assemblies upon which they act. The active surface and alignment system stabilizes the optical path against misalignment (of optical elements) and local deformations (M1 surface). The flexible body control uses separate sensor circuits to compensate deformations of structural parts outside (or below) the optical system (e.g., thermal deformations of the azimuth structure and the rocking chair, and changing unevenness of the azimuth track over time).}

\begin{figure}[tbh]
        \centering
        \includegraphics[width =\hsize, trim={0.2cm 0cm 0.2cm 0cm},clip]{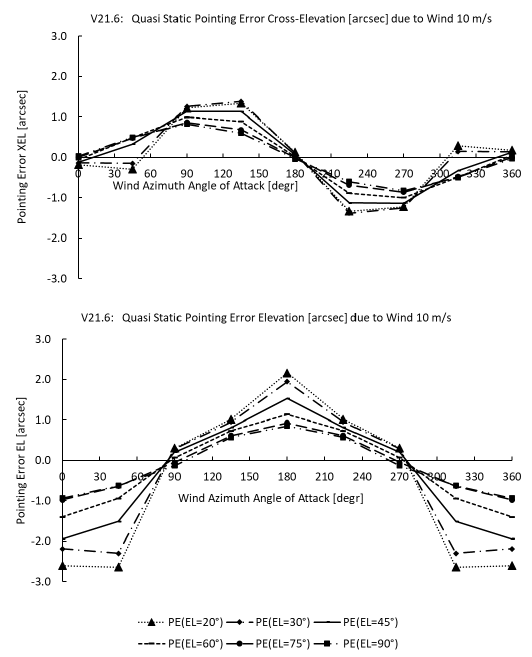}
        \caption{Pointing errors in cross-elevation (XEL, upper panel) and elevation (EL, lower panel) due to $10~\rm m \, s^{-1}$ static wind for different telescope elevation pointings and relative azimuthal angles for the wind. The diagram shows elevation and cross-elevation pointing errors separately (cross-elevation is the axis of rotation perpendicular to the elevation axis, the position of which is defined by the current elevation angle). The abscissa shows the wind azimuth angle of attack. The ordinate indicates the directional error in arcseconds. Each line represents the pointing error for a specific elevation angle of the telescope.}
        \label{fig:perf_pe_wind}
\end{figure}

The most critical and difficult pointing errors are likely to come from wind loads. 
The wind loads can be divided into stationary and dynamic loads by the eigenfrequencies of the principal axis control loops, which are limited by the first eigenfrequencies of the structure. Figure~\ref{fig:perf_pe_wind} shows the residual pointing error due to large stationary wind loads (assuming that the measured values of the principal axis angle sensors are already compensated to zero).

\subsection{Metrology concepts}\label{sec:metrology_concepts}
\subsubsection{Overview of active primary surface approaches}

No metrology system that satisfies AtLAST's HWFE specification has been demonstrated yet in the field, and as such, this will be a key area for development in the next phase of the AtLAST project.  
Here, we instead discuss potential approaches for achieving our surface accuracy, drawing from the pathfinding works of others. 

We note that the approaches considered here are broadly and more properly termed ``active optics'' rather than ``adaptive optics'' because they correct deformation errors of the telescope as an instrument but do not correct for atmospheric phase variations in the way optical/IR adaptive optics do.
For AtLAST, the errors to be actively corrected to achieve the HWFE requirement from the aperture to the instruments are: the primary reflector surface deformations, the telescope focus, and the relative alignment of the optical components including the instruments.
Furthermore, it must be emphasized that the term ``active surface'' in this context is meant to go beyond the accuracy of an open-loop system where the repeatable gravity deformations can be corrected via look-up tables obtained from dedicated calibration campaigns.
For AtLAST, the aim is to actively measure the primary surface deformations to compensate continuously for slow transient effects from temperature changes and wind loads, averaging over 10 second intervals. 
A by-product of the live metrology will be lower beam uncertainty and better calibration accuracy.  
Calibration systematics have been one of the primary limitations or uncertainties in many mm-wave surveys, presenting few percent level systematics that affect constraints from, for example, CMB analyses \citep{Hasselfield2013, Lungu2022}.

AtLAST's primary reflector surface accuracy requirements (nighttime value of 20~$\mu$m RMS, as in Table~\ref{tab:design})  will require an active surface coupled with a metrology system to measure it continuously.
A number of large single-dish telescopes with active surfaces, such as the 100-meter GBT, have met or even exceeded their original surface accuracy requirements by relying on out-of-focus (or phase-retrieval) holography using bright, compact astronomical sources \citep{Hunter2011, Mason2014, White2022}.  
However, astro-holographic approaches also incur large observing overheads and can lead to unknown calibration systematics as the beam quality degrades between focusing procedures.  
Recently, many of the large single-dish telescopes operating at mm wavelengths have begun to upgrade their systems to implement metrology,
{providing} a valuable pathfinder. AtLAST plans from the beginning to include in its design the ability to have closed-loop metrology continuously updating the surface corrections within seconds.

To our knowledge, there are three main approaches\footnote{Additionally, live photogrammetry, updating on timescales of a few seconds, can be expected to achieve 150-500~$\mu$m accuracy depending on the configuration of targets on M1 and the angles covered by the cameras \citep[see][for simulations of different configurations on the 64-m SRT]{Buffa2016}.} now being developed into potentially viable metrology systems: those based on direct measurements using lasers to scan the surface \citep[e.g.,][]{Salas2020}, those based on millimetric wavefront sensing \citep[e.g.,][]{Naylor2014, Tamura2020}, and those using laser interferometry to track changes in the path length \citep{Rakich2016, Attoli2023}. 
The first was being developed on the 100-m Green Bank Telescope as the Laser Antenna Surface Scanning Instrument but appears to be limited by the scanning strategy to slower-than-real-time performance and will require further technical development before a full demonstration (private communications with GBT staff).
The second technology has now been tested as a two-element prototype on the Nobeyama 45-meter telescope \citep{Nakano2022}, and has demonstrated a precision of $8~\mu$m, though further work will be required to fully implement it and show how it scales.
The third technology, being developed for the Giant Magellan Telescope (GMT) and the 64-m SRT, relies on the Etalon Absolute Multiline Technology$^\textsc{TM,}$\footnote{See \href{https://www.etalonproducts.com/en/products/absolute-multiline-technology/}{https://www.etalonproducts.com/en/products/absolute-multiline-technology/}.} to achieve, in principle, a precision of $0.5~\mu$m, satisfying our engineering partner's (OHB's) internal engineering design principle, which is that any measurement system used in a control system should deliver $\geq 10\times$ higher precision than the surface accuracy requirement in order to implement real-time corrections.

\begin{figure*}[tbh]
        \centering
        \includegraphics[width =0.9\hsize]{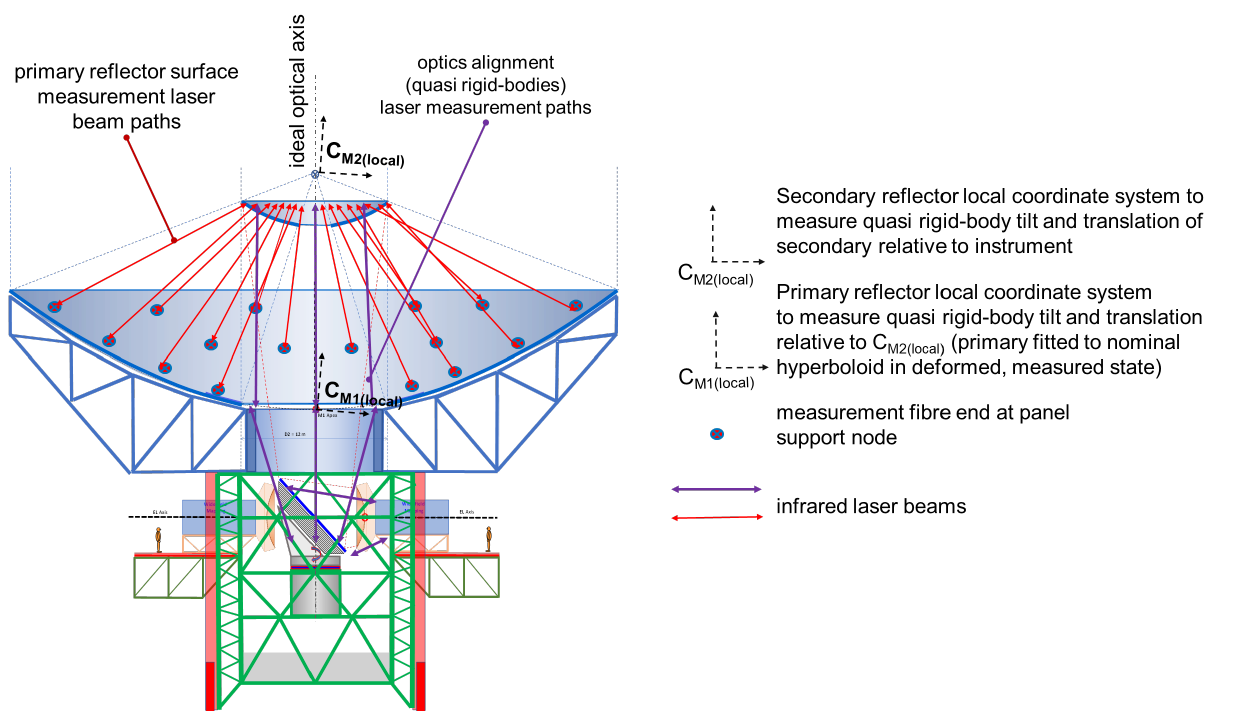}
        \caption{Sensor scheme for optical alignment and primary mirror surface deformation to ensure the optical quality by using Etalon Absolute Multiline Technology$^\textsc{TM}$ for absolute distance measurements.}
        \label{fig:metrology_act_surf}
\end{figure*}

\begin{figure*}[tbh]
        \centering
        \includegraphics[width =0.99\hsize, trim={0cm 0cm 0cm 0cm},clip]{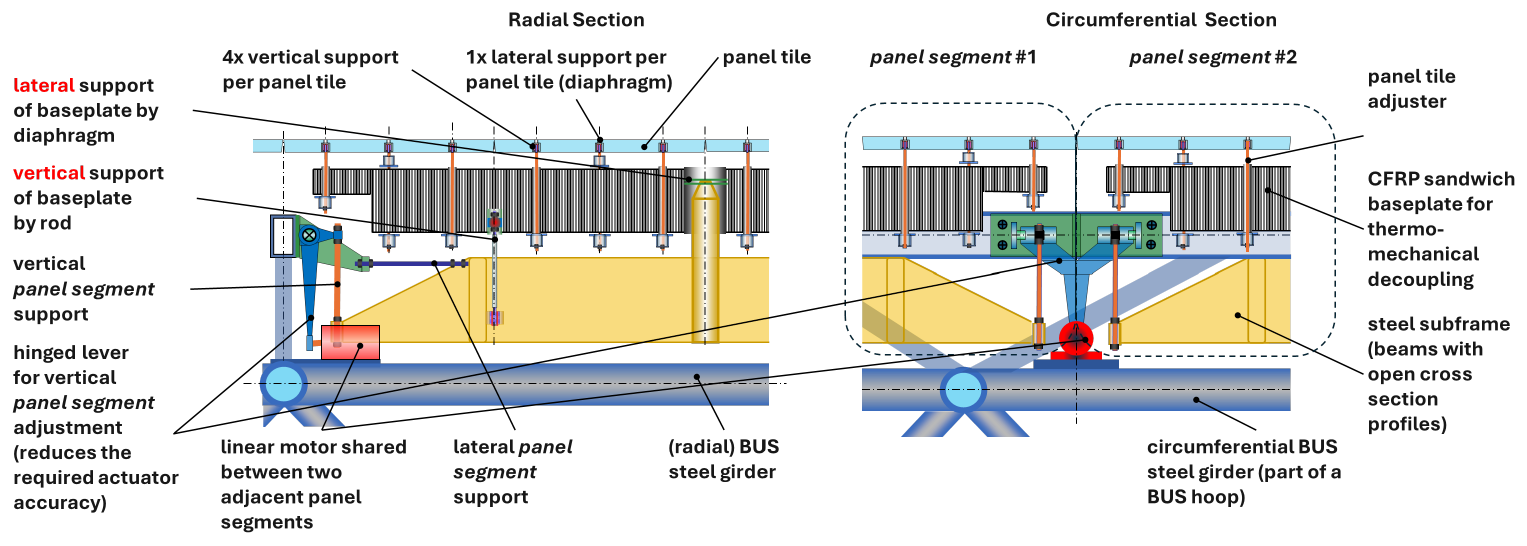}
        \caption{Arrangement of linear actuators at the corners of panel segments. The left panel shows the radial section (tangential view) of the panel segment motorized adjuster, while the right panel shows the circumferential section (radial view).}\label{fig:metrology_act_surf_motors}
\end{figure*}

\begin{figure*}[tbh]
        \centering
        \includegraphics[width =0.95\hsize]{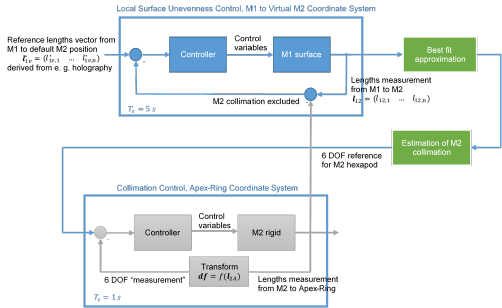}
        \caption{Main reflector panel control loop, consisting of active optics (active, flexible surface) and alignment of optical elements in rigid body DOFs. Here, the term ``apex-ring'' refers to the opening in M1 where several of the Etalon metrology systems will be mounted (see the beam paths represented by purple lines in Fig.~\ref{fig:metrology_act_surf}).
        }\label{fig:metrology_wfe_ctrl_loop}
\end{figure*}

\subsubsection{AtLAST active surface and optical alignment approach}\label{sec:ActiveOptics_and_Alignment}

As was introduced in the previous {section}, a system that ensures sufficient optical quality can be subdivided in two major functional areas:
\begin{itemize}
\item M1 reflector surface local deformations {after the subtraction of six DOFs fit to the nominal shape. The overall uncorrected HWFE at night is found in the finite element modeling to be $\lesssim 210~\mu$m (the quadrature sum of the components in Table~\ref{tab:surferrors}). Based on prior experience, we expect to reduce this $\lesssim 20~\mu$m through the use of an active surface with closed-loop metrology.}
\item M2, M3 and instrument misalignment in six DOFs due to deformation of their supporting structure.
\end{itemize}
The major deformations to be compensated and their contribution in HWFE based on FEA are listed in Table~\ref{tab:surferrors}. We note that these are only a subset of the error contributors to the overall half wavefront error. However, it can be seen that the closed-loop approach will be required in order to  compensate slow transient deformations due to temperature changes. In the \href{https://zenodo.org/records/14673697}{online supplement}, it can be seen that there is one case that cannot be compensated, which is when a temperature difference $\Delta T=1^\circ$C between the M1 backup structure and the rocking chair is present.  This will contribute a HWFE of 10~$\mu$m~RMS. 

The control loop can be subdivided into the following major components:
\begin{enumerate}
    \item Sensors detecting deformations of the primary reflector surface relative to the secondary reflector, where the secondary reflector is considered as a virtual rigid body (e.g.,\ by fitting its deformation to its nominal shape). The choice of a sensor system that achieves the required accuracy over the large, absolute dimensions of the primary reflector while being exposed to the environment is a major challenge. The only sensor system showing both the required accuracy and technology readiness level (TRL)\footnote{See \href{https://horizoneuropencpportal.eu/store/trl-assessment}{https://horizoneuropencpportal.eu/store/trl-assessment} for definitions of TRL.} so far is the Etalon Absolute Multiline Technology$^\textsc{TM}$.  Its TRL is estimated to be 5-6, and thus will require further development and testing before being considered ready for AtLAST.
    \item Sensors detecting alignment changes between the optical elements. Here, the same technology as for the primary reflector surface can be used. The Etalon Absolute Multiline Technology$^\textsc{TM}$ is already used for this purpose in optical telescopes such as ESO's Very Large Telescope (VLT), the Large Binocular Telescope (LBT) and the future Giant Magellan Telescope \citep[GMT;][]{Rakich2016}.\footnote{See also \href{https://www.etalonproducts.com/en/absolute-multiline-technology-for-the-giant-magelan-telescope/}{https://www.etalonproducts.com/en/absolute-multiline-technology-for-the-giant-magelan-telescope/}.}
    The arrangement of the sensors (surface and alignment) can be seen in Fig.~\ref{fig:metrology_act_surf}.
    \item Linear actuators for the alignment of the primary reflector panel segments. It can be expected that not all panel segment corners need to be supported by actuators since the deformation patterns to be corrected originate from the backup structure, and are of large scale, corresponding to or including mainly low order Zernike terms. A scheme for how the linear actuators can be arranged at the corners of a panel segment frame is shown in Fig.~\ref{fig:metrology_act_surf_motors}.
    \item Actuators controlling six (or fewer if not required) rigid body DOFs of the optical elements (except for the primary reflector). The secondary reflector and optionally the instrument are intended to be adjustable by six linear actuators comparable to a Stewart platform but with a different geometrical arrangement of the actuators.
    \item The control system can be considered as a cascaded design.
    The primary, inner, faster ($T_s=1$~s) control loop is the one that keeps the optical elements aligned and the secondary, outer, slower loop ($T_s=10$~s) keeps the M1 surface in its nominal {shape} as needed to meet the half wavefront error requirements from the aperture to the instruments.
    A sketch can be seen in Fig.~\ref{fig:metrology_wfe_ctrl_loop}.
\end{enumerate}

\subsubsection{Closed-loop pointing correction: Flexible body compensation}\label{sec:flexible_body_compensation}

The pointing direction of the overall optical assembly can be corrected using solely the main axis drives for corrections. Temperature and distance measurement sensors in the main support structure will be used to detect relative deformations within the rocking chair and the azimuth rotating structure.
The corrections by the active alignment system described in the previous section couple with the pointing direction and thus will be considered by the FBC system. {The errors in the pointing direction resulting from the optical alignment movements} will be compensated by the main axes drives as necessary.

\subsection{Panel surfacing and solar scattering}\label{sec:solar}
 
The main challenge to performing solar observations with a (sub)millimeter telescope is to reduce the loading power from the Sun that, if focused onto the receiver or any other component, would likely be dangerous and would at minimum jeopardize equipment, while still retaining the ability to carry out the observations.
Due to this challenge and the associated risks, most millimeter and submillimeter telescopes cannot observe the Sun; ALMA is a notable exception and can provide an example for AtLAST in this regard. 

In fact, the ALMA observatory features several different antenna designs, providing three independent, demonstrated approaches for implementing surface treatments that would enable direct observations of the Sun. 
All of these approaches rely on some form of scattering of the infrared and optical radiation by adding features that are sub-wavelength sub-terahertz frequencies, and with surface roughnesses well below the 20~$\mu$m RMS half wavefront error demanded by AtLAST's science requirements (Table~\ref{tab:design}).

In short, the approaches in use by ALMA are 1) sandblasting of the mandrel used to electro-form the nickel subreflector panels, which are then plated in aluminum to avoid corrosion, 2) chemical etching of aluminum subreflector panels, or 3) scalloping of the subreflector surfaces during their manufacture by milling (direct machining).  Control of the feature size and shape in the third case is achieved through the choice of tool size and milling feed rate parameters.
 
ALMA Memo \#329 \citep{alma_memo_329}, which supersedes ALMA Memo \#266 \citep{alma_memo_266}, shows that the efficiency that the panel reflects solar radiation at frequencies well above AtLAST's range of interest can be reduced to only $\eta \approx 0.05$ through the use of simple triangular grooves. Later, \cite{alma_memo_575} showed that a more random, chemically etched surface will have a specular reflection $\eta \approx 0.013$, roughly three times lower. 

We further note an advantage AtLAST has in this regard, due to its 12-meter large secondary mirror (M2).  While the collecting area of AtLAST M1 is larger than that of an ALMA antenna, the ratio of the areas of the primary to the secondary reflectors is much smaller for AtLAST than for ALMA, meaning the energy reflected by the AtLAST primary mirror will be distributed over a greater area. 
Concretely, that ratio is $(50/12)^2 = 17.4$ for AtLAST, and $(12/0.75)^2 = 256$ for an ALMA 12-m element, which features a 0.75 meter diameter secondary mirror.
Considering the intrinsic concentration of scattered light at the AtLAST secondary will be $\sim 14.7\times$ lower than it is for ALMA, this will further reduce the heat loading and therefore reduce the challenges associated with implementing solar observing modes. 


\section{Instrumentation considerations}\label{sec:insts}

AtLAST is designed to be a facility telescope serving a broad range of scientific goals including primary investigator proposed science and surveys. 
Following the recommendations of the instrumentation community and of the operations study, we have designed the telescope to facilitate the large number of science goals described in \cite{Ramasawmy2022}, \cite{Booth2024a}, and the AtLAST science cases \citep{Akiyama2023, Cordiner2024, DiMascolo2024, Klaassen2024, Lee2024, Liu2024, Orlowski-Scherer2024, vanKampen2024, Wedemeyer2024}. The design allows up to six nominal dedicated instrument positions, selected only by an axial rotation of the tertiary mirror.  As is discussed in \cite{Groppi2019},  \cite{AtLAST_memo_3}, and \cite{Kohno2020}, the types of instrumentation are expected to fall into one of the four following broad categories:

\begin{itemize}
    \item Multi-chroic or multiband imaging polarimeters, like those on SO \citep{Zhu2021, Bhandarkar2022}, CCAT \citep{Vavagiakis2018}, and CMB-S4 \citep{Gallardo2022}.
    \item Wideband direct-detection spectrometers like DESHIMA \citep{Taniguchi2022} or in the EXperiment for Cryogenic Large-Aperture Intensity Mapping \citep[EXCLAIM;][]{Volpert2022}.
    \item Heterodyne focal plane arrays like those in the CCAT Heterodyne Array Instrument \citep[CHAI;][]{Stacey2018} or novel integrated approaches like those in \cite{Shan2020} and \cite{Wenninger2023}.
    \item Multi-band single-beam heterodyne receivers that would allow AtLAST to participate in VLBI or, potentially, the Next-Generation Event Horizon Telescope (ngEHT) observational campaigns \citep{Johnson2023}.
\end{itemize}

Of these, the power and cryogenic cooling requirements will likely be driven by the first class of item in the above list, as those tend to be the most massive.

\subsection{Instrument sizes and cooling requirements}\label{sec:inst_sizes}

\begin{figure}[tbh]
    \centering
    \includegraphics[width=\hsize, trim={55mm 5mm 75mm 5mm},clip]{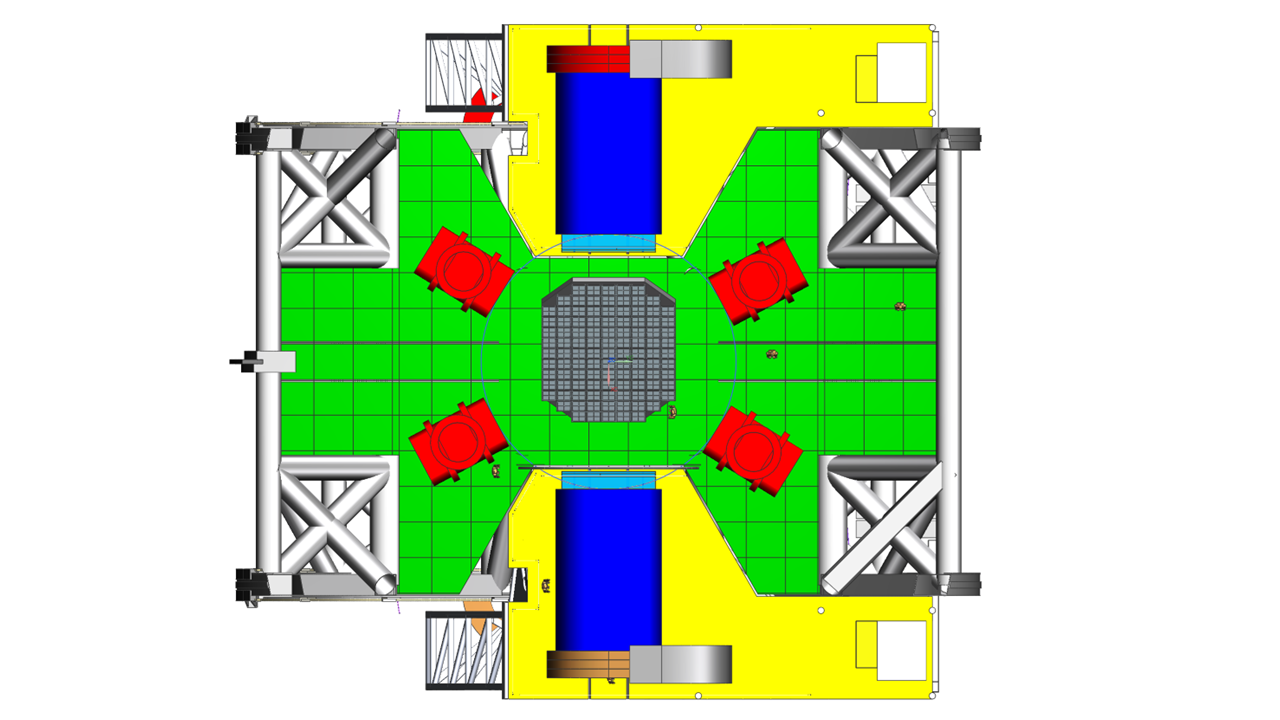}
    \caption{Top view of the AtLAST receiver cabin. The Cassegrain-mounted instruments are shown in red. The access platform for these instruments, which moves with the elevation rotating structure, is shown in green.  The Nasmyth-mounted instruments (blue cylinders) are located on the yellow platforms, which do not move with the elevation rotating structure. 
    Each Nasmyth platform has an area of 154~m$^2$. The receiver cabin area for the four Cassegrain instruments has a size of $\sim 410~\rm m^2$, while the floor to ceiling height is $\approx6.3$~meters.  Top views of average adult humans are shown for scale.}
    \label{fig:receiver_cabin}
\end{figure}

It is difficult to convey intuitively just how large each of these instruments can ultimately be in comparison to current facility instruments.
{For instance, the four smaller Cassegrain-mounted instruments each have a diameter of $\sim 2.6$ meters, which is about the width of a standard shipping container.} These are specified to be up to 10 tons each, and will be secured to the top of the elevation wheel structure as they are co-moving in elevation (see the red instruments in Fig.~\ref{fig:receiver_cabin}).  
They are expected to be comparable in diameter (and focal ratio, as is discussed in Sect.~\ref{sec:optics}) to the receiver designs for several other facilities (e.g.,\ SO, CCAT, and CMB-S4).
Based on this experience, the power requirements for the cryogenic cooling systems for the Cassegrain instruments are estimated to be $\leq 40$~kW per receiver (Table~\ref{tab:energy}). For example, \cite{Zhu2021} noted that the SO Large Aperture Telescope Receiver (LATR) uses two Cryomech$^\textsc{TM}$ PT90\footnote{See \href{https://bluefors.com/products/pulse-tube-cryocoolers/pt90/}{https://bluefors.com/products/pulse-tube-cryocoolers/pt90/}.} and two Cryomech$^\textsc{TM}$ PT420\footnote{See \href{https://bluefors.com/products/pulse-tube-cryocoolers/pt420/}{https://bluefors.com/products/pulse-tube-cryocoolers/pt420/}.} pulse tubes, which total $\approx 34$~kW.
While tilting and tipping of the Cassegrain instruments could present an engineering challenge for the cryogenics, a number of instruments have been operating while tipping over a greater range of angles for over a decade \citep[e.g.,][]{Dicker2008, Tan2008, Dicker2014}.  Furthermore, dedicated tests \citep{Tsan2021} have shown that the impacts of tipping a cryogenic pulse tube typically used for sub-4~K receivers as much as $\pm 35^\circ$ are minimal.
We also note that the heterodyne receivers in ALMA are co-moving in elevation and have been operating for over a decade.

The two larger instruments, meanwhile, are $4\times$ larger as they are intended to take advantage of the full 4.7-meter focal surface of AtLAST, corresponding to the full 2$^\circ$ diameter geometric FoV.
We chose to design the mount points for these to be Nasmyth (i.e.,\ fixed in elevation while the telescope points; see the blue instruments in Fig.~\ref{fig:receiver_cabin}). This is due to their exceptional volumes and masses (up to 30 tons, see Table~\ref{tab:masses}), the desire to facilitate operations, installation, access to the large instruments, cabling, and electronics for maintenance, and to accommodate the likely scenario they will be modular.
As the focal surface areas of the Nasmyth instruments are four times larger than those of the smaller instruments, we estimate the cooling requirements to be four times larger as well, resulting in a power budget of 160~kW per instrument for the cryogenics alone (Table~\ref{tab:energy}).

All six instruments are allocated a length of up to 6 meters, about the length of a typical shipping container used for freight, which will allow for readout and back-end electronics to be directly mounted on the backs of the instruments or the floor nearby.  
We note that the Nasmyth support platforms put the weight of the instruments directly on the azimuthal support structure, rather than the primary mirror backup structure.
This places an additional demand on the antenna stiffness, as the Nasmyth instruments are no longer directly attached to the same support structure that carries the telescope optics (i.e., M1, M2, and M3).  
The FEA shows that the relative pointing offset between the optical FoV and the instrument is acceptable, with the worst offsets during azimuthal turnaround (see Table~\ref{tab:eigenfreq}, modes \#2 and \#4).

\subsection{Receiver cabin and instrument installation}\label{sec:rxcabin}

\begin{figure*}[tbh]
    \centering
    \includegraphics[width=0.95\hsize, trim={0mm 30mm 83mm 2mm},clip]{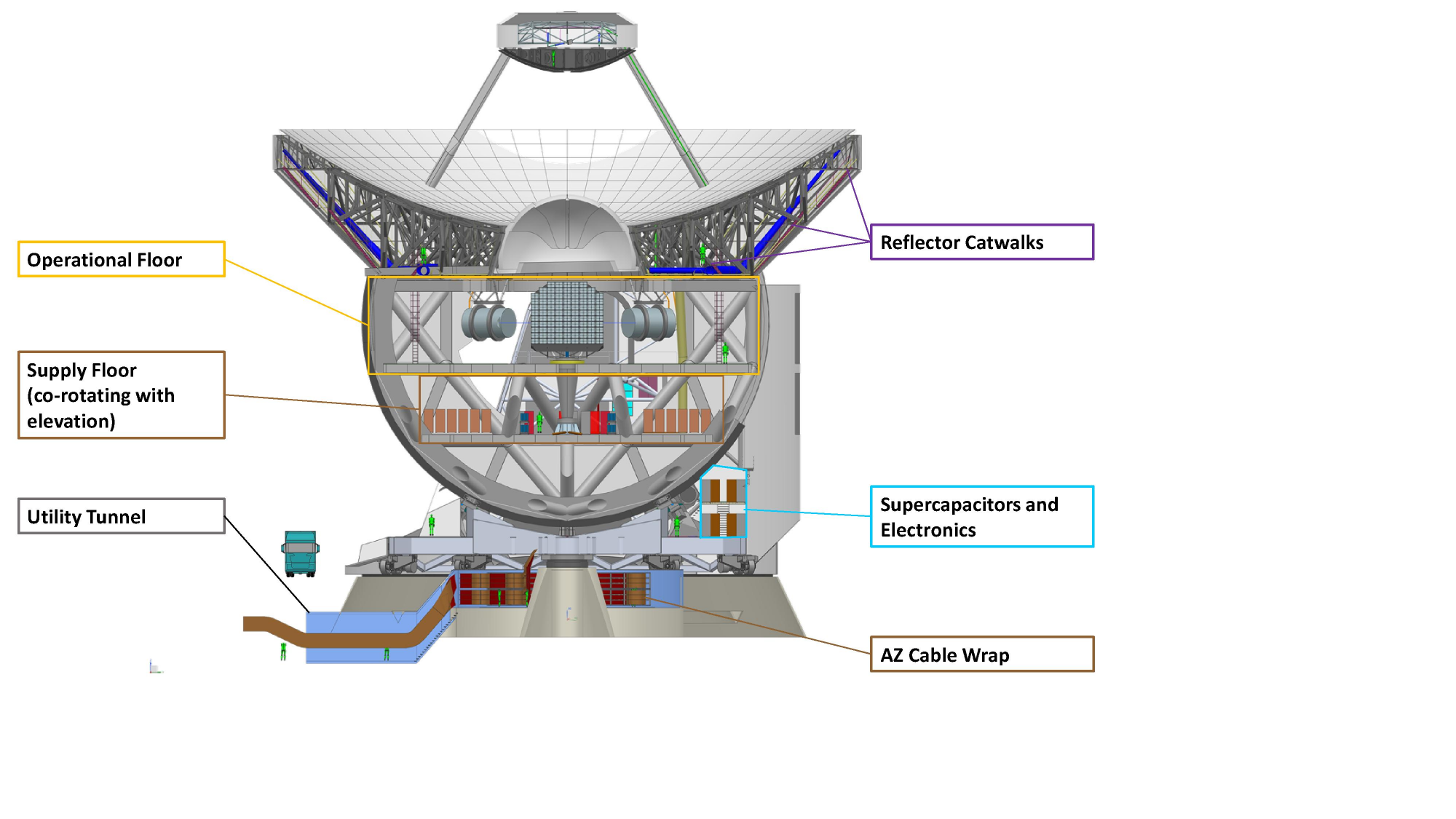}
    \caption{Section cutaway overview of the operational areas within the AtLAST telescope. The operational floor, outlined in yellow, highlights the receiver cabin volume.  The supply floor, outlined in brown, shows one of the areas reserved for computing equipment, compressors, and other equipment that can corotate in elevation.  The blue section in the lower right denotes where the supercapacitors for the regenerative braking system will be housed (Sect.~\ref{sec:energy}).}
    \label{fig:CAD_cutaway}
\end{figure*}

A cutaway view of the AtLAST CAD model is shown in Fig.~\ref{fig:CAD_cutaway}, revealing the telescope interior.  At the top, a catwalk structure allows access to the backs of the main reflectors' panels, the backup structure's heating, ventilation, and air conditioning (HVAC) system (blue), and the tile segment actuators. A yellow box in the same figure shows the main floor of the receiver cabin, where the science instruments themselves are located. 
Along with the instrument position within the telescope comes the task to install the six instruments into their operational position. This is essential during commissioning, and is also expected be required occasionally during the lifetime of the instrument (e.g., for maintenance and upgrades). Taking the foundation of the AZ-track as the zero-level, the instrument must be lifted by 18.6~m to reach the operational floor.

\begin{figure}[tbh]
    \centering
    \includegraphics[width=\hsize, trim={0mm 4mm 0mm 12mm},clip]{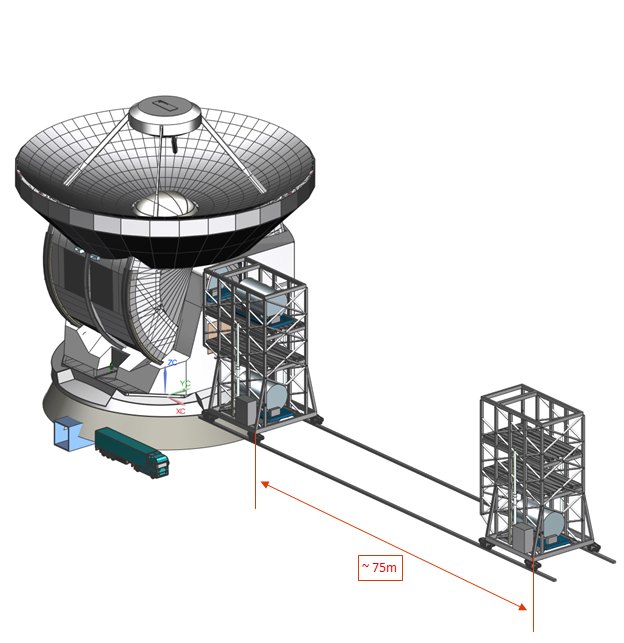}
    \caption{Design concept for the instrument installation tower. A pair of rails allows the tower to be moved toward and away from the telescope, while the instruments are loaded onto their respective platforms from the top position of the tower elevator.}
    \label{fig:inst_tower}
\end{figure}

We identified a system completely separate from the telescope structure to fit best AtLAST's requirements (shown in Fig.~\ref{fig:inst_tower}). A stand-alone system can access all four sides of the telescope, so that all instruments (and as necessary other devices) can be lifted with the same shared device and the transport systems in the telescope can be reduced to a minimum. As a result, it removes weight from the telescope structure and thus reduces the main axis moments of inertia and system complexity. As an additional benefit, this lifting device is also to be equipped so that it can be used for M2 maintenance activities.
Since such a separate system would interfere with the beam, it must either be foldable or be able to be removed from the vicinity of the telescope. It was determined that the minimum distance must be at least 75~m to be clear of the beam when the telescope is pointing at its lower elevation limit. The rail mounted system shown in Fig.~\ref{fig:inst_tower} was designed by taking into account the prospective AtLAST sites,\footnote{See \href{https://atlast-telescope.org/news-and-events/news/2023/measuring-the-wind-at-5000-m-above-sea-level.html}{``Measuring the wind at 5000 m above sea level'' (2023)}.} both of which are rather flat. For transportation, the instruments should be mounted on special, actively driven racks (Fig.~\ref{fig:inst_tower}, blue platform below the science instrument), which can also remain on the instruments for transportation to the final maintenance building using a customized trailer. In addition to this instrument lifting device, there are two standard passenger elevators located at each Nasmyth tower for transportation of personnel and small equipment.

Below the receiver cabin in Fig.~\ref{fig:CAD_cutaway}, the brown boxed region shows the {supply floor}. This space is dedicated for support electronics, computing or data storage, and auxiliary equipment for the Cassegrain instruments and the active M1 surface system. 
Vibrations from compressors and other equipment that must be located on the supply floor will be mitigated using vibration isolation and damping
While the precise amount of cooling power needed for the supply floor is not yet known, we budget up to 1~MW of power in Table~\ref{tab:energy} for the electronics and compressors.  In order to avoid dissipating their heat loads internally within the receiver cabin, the design includes the use of cooling fluid to carry the heat to the outside environment, where the average temperature is $\sim 0^\circ$C \citep[e.g.,][]{Morris2024}.

Below the supply floor, the light blue region in Fig.~\ref{fig:CAD_cutaway} shows the space on the azimuth structure hosting the supercapacitors to be used in the regenerative braking (or energy recovery) system and the main axis control systems (see Sect.~\ref{sec:energy} and \citealt{Kiselev2024}).

\begin{figure}[tbh]
    \centering
    \includegraphics[width=\hsize, trim={30mm 10mm 40mm 10mm},clip]{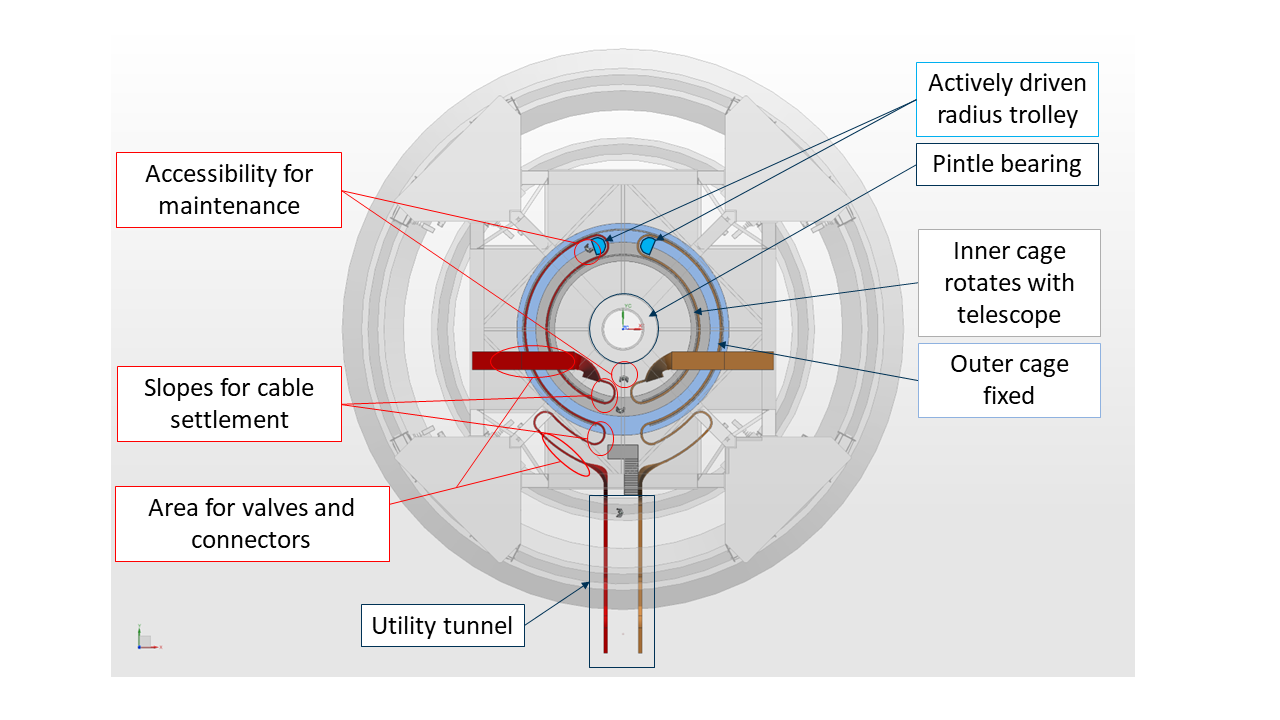}
    \caption{Azimuth cable wrap layout, also partially visible as a side view in Fig.~\ref{fig:CAD_cutaway}.}
    \label{fig:az_wrap}
\end{figure}

To supply all these systems with sufficient electrical power, helium, data network, chilled
water and compressed air most of the area within the foundation -- below the azimuth wheel on track system -- is reserved for the azimuth cable wrap. Figure~\ref{fig:az_wrap} presents the minimum sizes needed for cable wrap in terms of bending radii and maintenance areas.
According to the telescope movement the azimuth cable wrap will have a turning range of 540$^\circ$
and must be able to deal with the {design goal of a} maximum speed of 3$^\circ \rm s^{-1}$ and maximum acceleration of 1$^\circ \rm s^{-2}$. 
Multiple drives and brakes are installed that pull the robotic energy chains---that is, the flexible structures within the cable wraps that protect and move with the cables and hoses---in a way that is synchronized to the telescope movement. To guide the chains and for protection of staff, the AtLAST cable wrap should have an inner and outer guide channel, with the inner one rotating with the telescope and cable wrap drives. The outer guide channel shall be fixed to the foundation and has a diameter of  14.5 meters.
The most challenging design drivers are the size, maintenance requirements (accessibility) and the bearing design of the cable wrap.
The AZ cable wrap is served by a utility tunnel that runs beneath the telescope’s azimuth rails (see Fig.~\ref{fig:CAD_cutaway} for comparison).

\begin{figure}[tbh]
    \centering
    \includegraphics[width=0.9\hsize]{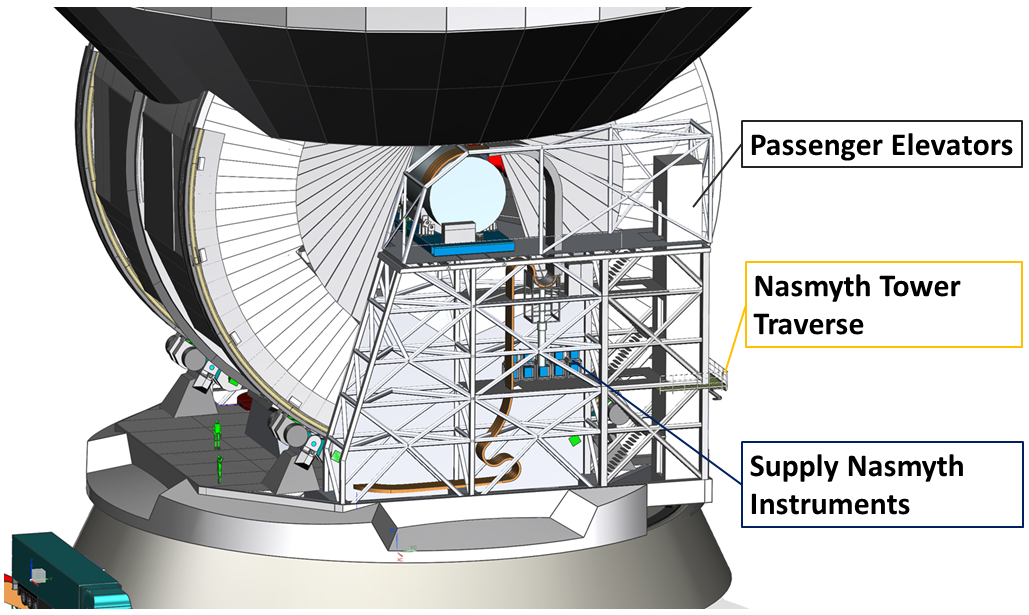}
    \caption{Nasmyth tower operational areas, showing how the antenna housing remains closed throughout changes in elevation. The passenger elevators and the Nasmyth tower traverse (stairs) are labeled as is shown, while the Nasmyth supply instruments indicated are compressors and electrical cabinets for the Nasmyth-mounted science instruments.
    }
    \label{fig:atlast_nasmyth}
\end{figure}

For systems that are corotating in elevation and that need to be supplied, for instance, the Cassegrain instruments or the active M1 surface, the elevation cable wraps are located in each Nasmyth tower, visible in Fig.~\ref{fig:atlast_nasmyth}. This guarantees excellent accessibility for maintenance.  The EL-cable wraps require a turning range of 76$^\circ$ (to create a safety margin beyond the nominal elevation range),  a maximum speed of 3$^\circ \rm s^{-1}$, and a maximum acceleration of 1$^\circ \rm s^{-2}$. They are also actively driven by linear actuators. 
The Nasmyth instruments can be supplied and supported separately with equipment hosted in the areas within the Nasmyth towers (see Fig.~\ref{fig:atlast_nasmyth}).

\begin{figure}
    \centering
    \includegraphics[width=\hsize, trim={0mm 0mm 0mm 0mm},clip]{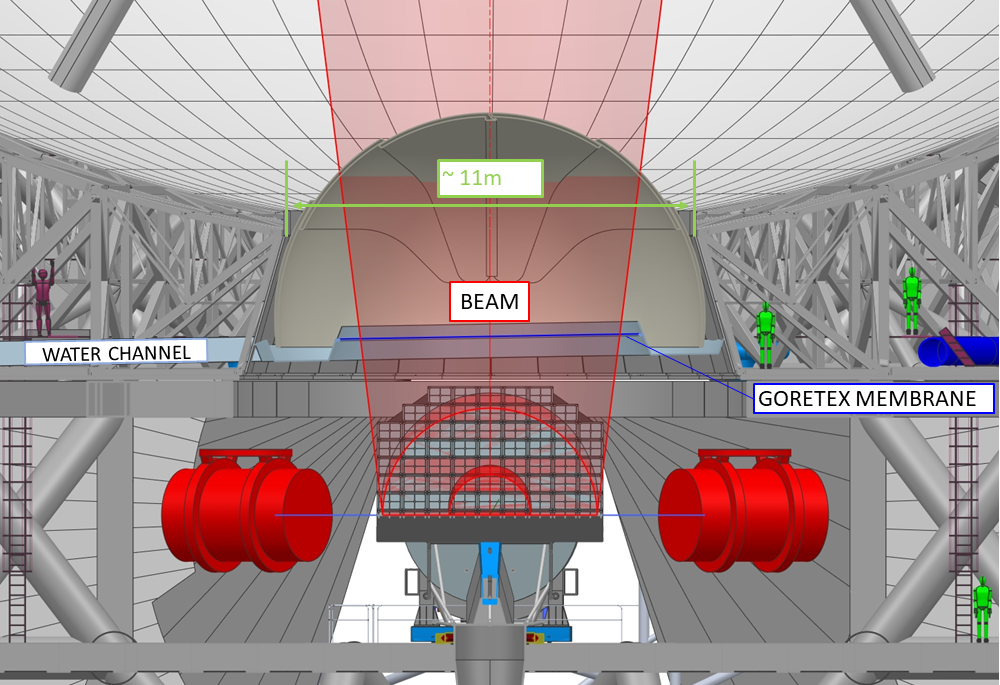}
    \caption{Section cutaway showing the receiver cabin protection by a spherical dome and the optically transparent Goretex\textsuperscript{\textregistered} membrane.}
    \label{fig:dome_membrane}
\end{figure}


Figure~\ref{fig:dome_membrane} shows the planned concept for protecting the receiver cabin from the elements (e.g., rain, snow, wind, and dust). Along with the optical layout described in Sect.~\ref{sec:optics} comes the fact that the M1 reflector has an opening of 11~meters in its middle. Given the environment in which AtLAST will be situated, it was decided to have two complementary systems for protecting the receiver cabin from the outside elements. As a baseline, the function to protect the interior during operation is to be executed by a Gore-Tex\textsuperscript{\textregistered} membrane that is considered acceptable for submillimeter wavelengths \citep[e.g.,][]{alma_memo_309}, and has been demonstrated at larger sizes by the JCMT, where the screen shields the entire 15-meter telescope \citep[e.g][]{Hills1990}. 
We note, however, that there is evidence that Gore-Tex\textsuperscript{\textregistered} may not be ideal in terms of high frequency transmission and polarization properties \citep{Greaves2003, Mairs2021}, and further investigation into alternative materials is needed.  

During the non-operating period and for extreme weather conditions a second, more robust system is foreseen. With regard to the areas required for the main reflector backup structure and the beam, we find that a spherical shape is best suited to AtLAST's conditions. Both a fiberglass shell widely used for small telescopes or a tent shell as planned should meet AtLAST’s requirements, with a fiberglass shell as baseline design \citep[as used by, e.g., the European Solar Telescope; see][]{Hammerschlag2010}. The shell of this shutter also provides water management to channel rainwater and snow from M1 away from the telescope interior. The inner side of the M1 backup structure will be sealed in the plane of the CFRP sandwich baseplate of the structure (see Fig.~\ref{fig:fe_model_panelseg}). The shutter guides the water to outlets on the side of the telescope.

\subsection{Wobbler and nutator considerations}\label{sec:wobbler}

Many single-beam instruments on smaller facilities have benefited from the ability to rapidly switch between an on-source and off-source position.  Typically this is achieved through a chopper or nutating secondary mirror, which ``throws'' the beam a few arcseconds off-source for approximately half the integration time. 
For a single beam in the ideal case, the instrument is modulated in a square wave, and the modulation occurs at a rate faster than the atmosphere evolves. Since 50\% of the time is spent on source, and 50\% is on a background sky position that is assumed to be free of significant astronomical signals, there is a noise penalty for both.  First, the signal is only measured half the time, and second the measurement of the off-position is just as noisy, so the effective on-source integration time is comparable to 25\% of the total.\footnote{Given the advances in focal plane array technology, and the fast scanning capabilities of the AtLAST design, a small focal plane array with a sufficient number of elements may obviate the need for a wobbler in most cases, getting around the on/off-source noise penalty by keeping a number of beams on source at all times, while the rest of the beams sample the background sky signal. Proper forecasting in upcoming works will explore this.}

The secondary and tertiary mirrors of AtLAST are too large to serve as nutators, and making them smaller would sacrifice most of the FoV. Detecting broad and faint spectral lines, such as those expected in emission from massive galactic-scale molecular outflows \citep{Cicone2014, Cicone2018} and from the circumgalactic medium (CGM) of galaxies \citep{Lee2024, Cicone2021}, require stable spectral baselines. For these kinds of science cases, it may be worth to sacrifice a portion of the FoV to apply the wobbler concept developed by R.~Hills and presented in AtLAST Memo \#2 \citep{AtLAST_memo_2}. This memo provides a contingency plan that avoids modulating either the massive secondary or tertiary mirror.
This wobbler concept, inspired by his work on the JCMT, relies on Crossed-Dragone relay optics to reimage and allow the installation of a small wobbler at the pupil.  
One of the advantages of this concept over traditional approaches is that this wobbler is in principle optically perfect:  since it modulates at the pupil, it is modulating an exact optical image of the primary mirror.  Furthermore, the optics could be housed in the receiver, meaning they 1) do not interfere mechanically or require modification to the antenna optics, and it would allow them to be cooled to cryogenic temperatures, which would minimize any additional loading (noise).


\section{Conclusions and future steps}\label{sec:conclusions}

The next decade of developments is poised to revolutionize submillimeter astronomy, with AtLAST having the potential to become becoming the flagship in this effort. 
In this work, we have presented the first conceptual design for AtLAST, and the results of finite element modeling for structural deformations due to wind, gravity, and thermal effects. Most importantly, we show that the demanding science requirements drive us to an innovative design quite unlike any existing millimeter or submillimeter facility.
Recent complementary works provide more complete details on the antenna structural design and technical flow-down from the observational capabilities requirements \citep{Reichert2024}, the optical design and corrections \citep{Gallardo2024}, the effects of panel gaps and scattering on the beam \citep{Puddu2024}, and a novel regenerative braking and energy recovery system that includes both power-saving and peak-power-mitigation features \citep{Kiselev2024}.
   
After further, anticipated development activities in the coming years, we expect that the conceptual design presented here will be able to meet our technical requirements and achieve AtLAST's broad scientific goals.
In this work, we have identified and discussed in depth the key technical requirements needed to achieve AtLAST's science goals, which in turn drive the most stringent aspects of the design. Namely, these are: the large FoV, the high surface accuracy, fast scanning and acceleration, and the need to deliver a sustainable and upgradeable facility that will serve a new generation of astronomers and remain relevant for the next several decades.

At the same time, while we have gained sufficient confidence in the feasibility of the conceptual design, a few key aspects of the design have been identified for further testing, prototyping, and demonstration before construction can begin.  These include a field demonstration of the closed-loop metrology to keep the overall optical surface accurate to better than 20 $\mu$m RMS half wavefront error, field demonstrations of the power generation and energy recovery concepts, computational fluid dynamics simulations to fully assess the impact of wind and turbulence on the pointing and focus, and a seismic hazard analysis to deliver detailed loads and design criteria for the particular sites of interest. 
These topics will be addressed in the next phase design study, which was recently selected for funding  by the European Commission through a new Horizon Europe research infrastructure grant.  This next phase will commence in early 2025.
Despite the amount of work that remains to be done, AtLAST is on track to complete a full design and start with construction preparations later this decade, once adequate funding has been secured. 

\section*{Data availability}

Supplementary materials showing the primary reflector's half wavefront errors and plots of the antenna's natural modes are \href{https://zenodo.org/records/14673697}{available on Zenodo}.

\begin{acknowledgements}
This project has received funding from the European Union’s Horizon 2020 research and innovation program under grant agreement No.\ 951815 (AtLAST).\footnote{See
\href{https://cordis.europa.eu/project/id/951815}{https://cordis.europa.eu/project/id/951815}.}
F.M.M.\ acknowledges the UCM Mar{\'i}a Zambrano programme of the Spanish Ministry of Universities funded by the Next Generation European Union and is also partly supported by grant RTI2018-096188-B-I00 funded by the Spanish Ministry of Science and Innovation/State Agency of Research MCIN/AEI/10.13039/501100011033.
We are grateful for the guidance and inspiration from Richard Hills throughout the early phases of the telescope design.
{And finally, we thank the referees for their constructive comments, which significantly improved this work.}
\end{acknowledgements}


\bibliographystyle{aa}
\bibliography{references}

\begin{appendix}

\onecolumn

\section{Acronyms and abbreviations}\label{sec:appendix_acroy}
We list the main acronyms and abbreviations used throughout this work in Table~\ref{tab:jargon}.  

\begin{table*}[]
  \caption{Acronyms and abbreviations used throughout the text.}
  \label{tab:jargon}
  \centering
  \begin{tabular}{l l}
    \hline\hline
    Acronym & Meaning \\
    \hline
    ACA  & Atacama Compact Array (Morita Array)\\
    ACT & Atacama Cosmology Telescope\\
    ALMA & Atacama Large Millimeter/Submillimeter Array\\
    APEX & Atacama Pathfinder EXperiment\\
    ARS  & azimuthal rotating structure\\
    AtLAST & Atacama Large Aperture Submm Telescope\\
    AZ   & azimuth\\
    BUS  & back-up structure\\
    CAD  & computer-aided design\\
    CFRP & carbon fiber reinforced plastics\\
    CHAI & CCAT Heterodyne Array Instrument\\
    CIB  & cosmic infrared background\\
    CL  & closed-loop (dynamic state space model)\\
    CMB  & cosmic microwave background\\
    CMB-S4 & Stage IV CMB experiment\\
    CSO  & Caltech Submillimeter Observatory\\
    DESHIMA & DEep Spectroscopic HIgh-redshift MApper\\
    DOF  & degree of freedom\\
    EL   & elevation\\
    ELT  & Extremely Large Telescope\\
    ERS  & elevation rotating structure\\
    ESO  & European Southern Observatory\\
    EXCLAIM & EXperiment for Cryogenic Large-Aperture Intensity Mapping\\
    FBC  & flexible body compensation\\
    FEA  & finite element analysis\\
    FEM  & finite element model\\
    FoM  & figure of merit\\
        FoV  & field of view\\
        FS   & focal surface (i.e., curved focal plane)\\
    FWHM & full width at half maximum\\
        FYST & Fred Young Submm Telescope (CCAT observatory)\\
    GBT  & Green Bank Telescope\\
    GMT & Giant Magellan Telescope\\
    GRASP & General Reflector Antenna Software Package\\
    HVAC & Heating, ventilation, and air conditioning\\
    HWFE & half wavefront error\\
    INAF & Istituto Nazionale di Astrofisica\\
    IPCC & Intergovernmental Panel on Climate Change\\
    IRAM & Institut de Radioastronomie Millimétrique\\
    JCMT & James Clerk Maxwell Telescope\\
    LBT & Large Binocular Telescope \\
    LMT/GTM & 50-meter Large Millimeter Telescope/Gran Telescopio Millimetrico {Alfonso Serrano} \\
    LOS & line of sight\\
    LUT  & Look-up table\\
    M1   & primary mirror\\
    M2   & secondary mirror\\
    M3   & tertiary mirror\\
    ngEHT & next generation Event Horizon Telescope\\
    ngVLA & next generation Very Large Array\\
    NOEMA & Northern Extended Millimeter Array\\
    PWV  & precipitable water vapor\\
    RMS  & root mean square\\
    SEST & Swedish-ESO Submillimeter Telescope \\
    SKA & Square Kilometer Array\\
        SO   & Simons Observatory\\
    SPEM & static pointing error (correction) model\\
    SRT & Sardinia Radio Telescope\\
    SZ   & Sunyaev-Zeldovich (effect)\\
    TMA & Three Mirror Anastigmatic design\\
    TRL & technology readiness level \\
    VLBI & Very Long Baseline Interferometry \\
    VLT & Very Large Telescope\\
    \hline
    \end{tabular}
\end{table*}

\end{appendix}

\end{document}